%% file: earlytype.tex
\documentclass[apj]{emulateapj}
\usepackage{apjfonts}
\usepackage{multirow}
\usepackage{color}
\usepackage{natbib}
\newcommand{\beq}{\begin{equation}}
\newcommand{\eeq}{\end{equation}}
\newcommand{\beqa}{\begin{eqnarray}}
\newcommand{\eeqa}{\end{eqnarray}}

\shorttitle{Star-Formation Bimodality in Early-Type Galaxies}
\shortauthors{Amblard et al.}

\begin{document}
\title{Star-Formation Bimodality in Early-Type Galaxies}
\author{
A. Amblard\altaffilmark{1,2},
L. Riguccini\altaffilmark{1,2},
P. Temi\altaffilmark{1},
S. Im\altaffilmark{1,2},
M. Fanelli\altaffilmark{1,2},
P. Serra\altaffilmark{3}
}


\altaffiltext{1}{NASA Ames Research Center, Moffett Field, CA, USA}
\altaffiltext{2}{BAER Institute, Sonoma, CA, USA}
\altaffiltext{3}{IAS, CNRS (UMR8617), Universit\'{e} Paris-Sud 11, B\^{a}timent 121, F-91400 Orsay, France}

\begin{abstract}
We compute the properties of a sample of 221 local early-type galaxies with
a spectral energy distribution (SED) modelling software, CIGALEMC. Concentrating
on the star forming activity and dust contents, we derive parameters such
as the specific star formation rate, the dust luminosity, dust mass and temperature.
52\% of our sample is composed of elliptical (E) galaxies and 48\% of lenticular (S0) 
galaxies. We find a larger proportion of S0 galaxies among galaxies with 
a large specific star formation rate (sSFR) and large specific dust emission.
The stronger activity of S0 galaxies is confirmed by larger dust masses. 
We investigate the relative proportion of active galactic nuclei (AGN) and star-forming 
(SF) galaxies in our sample using spectroscopic SDSS data and near-infrared selection 
techniques, and find a larger proportion of AGN dominated galaxy
 in the S0 sample than the E one. This could corroborate a scenario where blue galaxies evolve into 
red ellipticals by passing through a S0 AGN active period while quenching 
its star formation. Finally, we find a good agreement comparing our estimates 
with color indicators.


\end{abstract}

\section{Introduction}
\label{sec:introduction}

Elliptical (E) and lenticular (S0) galaxies, early-type galaxies (ETGs), are
among the most massive galaxies today. 
Their poorly known formation is subject  to much debate. 

The two main scenarios of ETG formation are the monolithic and 
hierarchical models. In the monolithic view, ETGs assembled most of their mass
quite early on (z > 2-3) as they merge with smaller substructures, they are 
characterized by a strong early star formation (SF) and then evolve passively 
into galaxies we see today. In the hierarchical view, ETGs are formed by
mergers of galaxies either rich (``wet'' merger) or poor (``dry'' merger) in gas. 
In this picture, it is not clear which merging path(s) ETGs follow.

ETGs constitute the majority of the red galaxies ($\sim$ 75\% \citealt{Driver:06}) in 
the bimodal color distribution of galaxies, their stellar population already 
transitioned from blue to red as their star-formation ceased
\citep{Faber:07,Hughes:09}.
 Indeed the most massive (> 10$^{10}$ M$_\odot$) ETGs 
are red and with little star formation \citep{Temi:09b};
their stellar content formed at an early epoch 
\citep{Trager:00, Cimatti:04, Thomas:05, Temi:05a, Temi:05b} 
and passively evolved to their present form. \\
However, multi-frequency observations in recent years have 
revealed that on closer inspection many E galaxies and even 
more S0 galaxies do in fact contain dust and cold gas in 
amounts that cannot be ignored \citep{Young:11,Temi:07a, Temi:07b}, sufficient to generate appreciable 
star formation at rates as large as 
several $M_{\odot}$ yr$^{-1}$ \citep{Temi:09a, Temi:09b}. 

Recent ultra-violet (UV) observations of large 
samples of ETGs present clear evidence of current star-formation 
\citep{Yi:05,Schawinski:07a,Kaviraj:07,Kaviraj:08,Kaviraj:10, Salim:12, Fang:12,Barway:13}.
Using UV-optical colors, \cite{Kaviraj:07} show that at least $\sim$30\% 
of UV-selected early-type galaxies at z < 0.11 have evidence 
of recent star-formation within the last 1 Gyr. The contribution
 from old stars to the UV flux is however uncertain.

The remarkable diversity of nearby ETG galaxies is most clearly expressed
in infrared emission where many S0 galaxies exhibit 
a range of unusual properties 
not often found in ellipticals.
\cite{Temi:09a, Temi:09b} showed a
banana-shaped 
mid (24$\mu$m) and far (70$\mu$m)  correlation containing only early-type galaxies.
While S0 galaxies are spread along
the entire correlation, most elliptical galaxies, for which star formation is negligeable,
occupy an extended region with a nearly flat $\log(L_{24}/L_K) \approx 30.2$. 
S0 galaxies have a star-forming diversity that mimics 
infrared colors of galaxies of any type. The SAURON survey \citep{Bacon:02,Dezeeuw:02} found that E and S0 can be separated into slow and fast rotators, the slow rotator group being composed mostly of ellipticals \citep{Cappellari:07,Emsellem:07}. Building in part on this finding, the $ATLAS^{\rm 3D}$ project \citep{Cappellari:11a} observed a sample of 260 local early-type galaxies (complete within 42 Mpc at -21.5 K band magnitude) in a large number of wavelengths, including optical integral-field spectroscopy (IFS). \cite{Emsellem:11} showed that most ETGs are fast rotators (86$\pm$2\%) and \cite{Cappellari:11b} showed that slow rotators are in general round ellipticals (E4 or rounder). They also showed that the slow/fast rotators classification improves the kinematic morphology-density T - $\Sigma$ relation over the E/S0 classification and argue that the kinematic classification may be more accurate than the morphologic classification which is plague by projection effects.

Lenticular S0 galaxies 
differ morphologically from E galaxies having 
stellar distributions in bulges and rotating 
disks that are somewhat less concentrated than E galaxies. 
In this paper, we investigate differences between Elliptical and Lenticular local
galaxy populations. 
Although the distinction between E and S0 morphologies is not always 
easy, many local S0 galaxies are quite distinct 
from ellipticals, having attributes consistent with recent star formation: 
significantly larger mid and far infrared luminosities, 
large masses of molecular gas, and 
younger stellar ages as inferred 
from optical absorption line spectra.
Since S0 galaxies at higher redshift are difficult to distinguish from 
E galaxies, particularly if the S0s are face-on, the two types of galaxies 
are often combined. Therefore
results from large surveys of ``early-type`` galaxies that combine 
S0 and E galaxies will be misleading if applied only to E.
However, the distribution of old population starlight in low redshift S0s, 
which are of interest here, define a unique and rather large fraction of early type galaxies.

As our main tool, we use a SED fitter, CIGALEMC \citep{Serra:11},
 capable of modeling our galaxy SED from UV to mm wavelength, which allows us to determine
the stellar mass, dust luminosity, stellar population age, bolometric luminosity 
and star formation rate of our galaxies in a less ambiguous manner than
single-band photometric indicators.\\ In the second part of the paper, we concentrate
on the far-infrared emission and estimate the dust mass and temperature of our
sample. We compare these quantities with the ones estimated from the full SED
and distinguish the two sub-population. In the third part, we investigate
the proportion of AGN in both the E and S0 population using SDSS spectroscopic
and NIR selection criteria. In the last part, we compare our analysis with optical, 
UV and Far-Infrared colors.

\section{The Data}
\label{sec:data}

We base our work on a sample of 225 Early-Types Galaxies (ETGs) from \citet{Temi:09b}. As pointed out 
in their work, a few sources present an uncertain morphological classification, in particular 
galaxies NGC 3656 and NGC 5666. We use the morphological type T 
from the HyperLeda database\footnote{http://leda.univ-lyon1.fr/} \citep{Paturel:03} and  exclude the 2 following sources: 
NGC3656 and NGC5666 because of the positive value of their T parameter,
 indicating that they are not ETGs as expected. Our sample is composed of 116 ellipticals 
(52\,\% of the sample), 35 E-S0 (16\,\%), 48 S0 (22\,\%) and 22 S0-a (10\,\%), which is roughly 
the same distribution as in \citet{Temi:09b}. In order to perform a reliable SED-fitting for 
each source, we need to have access to a large fraction of their SED from the UV to the FIR. To facilitate
the data analysis we concentrate primarly on 6 instruments which span theses wavelengths and have extensive
sky coverage : GALEX for the UV part, SDSS for the optical, 2MASS and IRAC-Spitzer for the near-Infrared (NIR) and mid-Infrared (MIR), and MIPS-Spitzer and IRAS for the mid and far-IR. We also use public data from Herschel-SPIRE\citep{Griffin:10} when available.\\
We address the representativeness of our sample using the wide (2 deg$^2$) equatorial COSMOS field \citep{Scoville:2007}. With a large statistic and multi wavelength datasets, and in particular an optical catalog with morphological indexes and a 24$\mu m$ sources catalog, it is a good source to answer this question.
We select 4\,605 ``local'' sources (z < 0.09, about D < 400 Mpc) in the COSMOS optical catalog\citep{Ilbert:2009}. 3\,873 of these sources are in the morphological catalog of \citet{Tasca:2009} (i.e. 84\%), and 994 are classified as ETGs (i.e. 26 \%)). 
In the 24 $\mu m$ catalog \citep{LeFloch:2009} 236 sources are lying at D<400Mpc, 206 are in the morphological catalog (i.e. 87\%), and 64 sources are classified as ETGs (i.e. ~ 31\%).
The percentage of 24$\mu m$ selected ETGs is slightly larger than the optical selected one, but it is not stastically significant, a 24$\mu m$ selection does not bias much an ETG sample.\\
We also compared the absolute B magnitude distribution of these three ETGs samples (COSMOS optical, COSMOS 24$\mu m$ and our sample). The comparison between the distribution of our sample of ETGs in absolute B magnitude with the one from the optically selected ETGs in COSMOS shows that they cover the same range of M$_B$ but the COSMOS distribution is slightly shifted towards brightest sources than the distribution of our sources, similarly to the COSMOS 24$\mu m$ sample. We conclude that our ETGs sample is quite representative of the overall distribution of ETGs in the local Universe.

\subsection{UV Observations}

To cover the UV part of the spectrum we use GALEX GR6 data release\footnote{http://galex.stsci.edu/GR6/}
, since it covers 25,000 deg$^2$ of the sky with a sensitivity down to m$_{AB}$=21 for the All Sky Imaging Survey (AIS) and m$_{AB}$=25 for the Deep Imaging Survey (DIS) \citep{Morrissey:07}. GALEX is a NASA satellite, equipped with two microchannel plate detectors imaging in the near-UV (NUV) at 2316 {\AA} and far-UV (FUV) at 1539 {\AA} \citep{Morrissey:07} and a grism to disperse light for low resolution spectroscopy.  The source position accuracy is about 0.34 arcseconds and the angular resolution of FUV and NUV is respectively 4.3 and 5.3 arcseconds. We find a match in GALEX catalog for 199 sources of our sample in the FUV band and 198 sources in the NUV band. We apply a galactic dust extinction correction, A(FUV)/E(B-V)=8.376 A(NUV)/E(B-V)=8.741, to GALEX data, assuming Milky Way dust with R$_v$=3.1 \citep{Cardelli:89,Marino:11}, and using NASA/IPAC Extragalactic Database (NED) E(B-V) values. The UV emission is a good indicator of the dust content and star formation rate of galaxies in our sample when compared with the optical data.

\subsection{Optical and NIR Observations}

We choose SDSS data to cover the optical range of the SED of our galaxies, given the large part of
the sky covered by SDSS (14,555 deg$^2$). The SDSS data have an angular resolution
of about 1.5 arcseconds.  We retrieve SDSS data through the Imaging Query Form interface\footnote{http://skyserver.sdss3.org/dr9/en/tools/search/IQS.asp}. 147 sources from our sample have an optical counterpart in the 5 bands of the Sloan Digital Sky Survey  (SDSS), u, g, r, i and z (respectively 0.335\, $\mu m$, 0.469\, 
$\mu m$, 0.616\, $\mu m$, 0.748\, $\mu m$ and 0.893\, $\mu m$).
At Near-Infrared wavelength, we use the extended source catalog of the Two Micron All Sky Survey (2MASS) data
, which contains 1,647,599 sources. 2MASS resolution is about 2 arcseconds and its source position accuracy
is about 0.5 arcseconds. The 10-$\sigma$ limit magnitude in the 3 filters J, H, K$_s$ is about 14.7, 13.9, 13.1
\footnote{http://www.ipac.caltech.edu/2mass/releases/allsky/doc/sec4\_5.html}.
220 Galaxies in our sample have counterparts in the 3 different filters, J, H and K$_s$ bands, 
at 1.24\, $\mu m$, 1.66\, $\mu m$ and 2.16\, $\mu m$ respectively. 

\subsection{InfraRed Observations}

The {\it Spitzer} Space Telescope provides data in the near-IR and mid-IR with the IRAC camera with 
4 channels imaging at 3.6, 4.5, 5.6 and 8\, $\mu m$ with about 2 arcsecond angular resolution 
and in the mid and far-IR with the MIPS instrument observing at 24, 70 and 160\,$\mu m$ at angular 
resolution of 6, 18 and 40 arcseconds \citep{Rieke:04}. We download data for our galaxies
from the NASA/IPAC Infrared Science Archive\footnote{http://sha.ipac.caltech.edu/applications/Spitzer/SHA}.
Reduction of the data follows \cite{Temi:09b} for MIPS and a similar treatment is applied on IRAC data to produce maps.
 The Mosaicing and Pointsource Extraction (MOPEX)\footnote{http://irsa.ipac.caltech.edu/data/SPITZER/docs/dataanalysistools/tools/\\mopex/mopexusersguide/} is used to process BCD data into corrected images and to co-add them into a mosaic.
To compute IRAC fluxes, a number of packages in the Image Reduction and Analysis Facility (IRAF)\footnote{http://iraf.noao.edu} 
are used for unit conversion, to remove artifacts, and to performe aperature photometry. We use
 the centroid sky fitting algorithm of the PHOT function (part of the APPHOT IRAF package) with an appropriate annulus and dannulu value and a constant photometric weighting scheme for wphot. 
For MIPS, flux densities are extracted from apertures that cover 
the entire optical disk (R25). Sky subtraction is performed by averaging values from multiple apertures placed around the target, avoiding any overlap with the faint extended emission from the galaxy.
Foreground stars and background galaxies present in the original mosaiced images are deleted before 
flux extraction is performed. These are identified by eye and cross-checked using surveys at other 
wavelengths (Digital Sky Survey and 2MASS). 
Fluxes of each IRAC and MIPS channel are obtained in mJy unit. We obtain a flux measurement
 for 165, 173, 164, 171, 204, 121, 93 of our galaxies at 3.6, 4.5, 5.6, 8, 24, 70, 160 $\mu$m respectively
\\
In order to have better constraints on the peak of the SED in the IR, to derive for example the dust 
temperature of our sources, we also include the observations from the {\it Infrared Astronomical Satellite} 
(IRAS). We obtain the IRAS dataset for our sample using the Scan Processing and 
Integration tool (Scanpi) which gives us the fluxes and errors in the 4 IRAS bands: 12, 25, 60 and 100 $\mu m$. 
The larger errors and the poor angular resolution (4 arcminutes at 100$\mu$m) of the IRAS observations compared 
with the rest of the available data encouraged us to check if the IRAS data agree well with the other data points. After removing the IRAS observations in case of conflict, we finally have a detection in the 12\, $\mu m$-band for 154 sources, in the 25\, $\mu m$-band for 124 sources, in the 60\, $\mu m$-band for 135 sources and in 
the 100\, $\mu m$-band for 141 sources.\\
The launch of the {\it Herschel}\footnote{Herschel is an ESA space observatory with science 
instruments provided by European-led Principal Investigator consortia and with important
 participation from NASA.} telescope allows unprecedented precisions at FIR wavelength.
We use public level2 data, which are maps of Astronomical Observation Requests (AORs), 
from the SPIRE instrument, downloaded from the Herschel Science 
Archive\footnote{http://herschel.esac.esa.int/Science\_Archive.shtml}. The SPIRE instrument
observes at 250, 350 and 500 $\mu m$ with an angular resolution of about 18, 25, 36 arcseconds \citep{Griffin:10}. 
Level2 maps are combined into a single map for each object using a simple pixel co-addition
technique (several AORs were requested for most objects). 
We subtract a background level from each image, the background is estimated 
by taking a median at 3 arcminutes around the source.
We compute the size of each galaxy on our 250 $\mu$m map by grouping neighboring pixels 
with a signal to noise 
ratio greater than 2 around the source location using a friend-of-friend algorithm.  
We then perform an aperture photometry of this angular size at each wavelength. 
The size of galaxies, which surface brightness is too low to perform this measurement (no pixels with a S/N larger than 2), is taken to be the major axis diameter 
from the NED. The aperture size is inspected by eyes to check for multiple sources or complex morphology. We obtained data for 75 galaxies and detected at the 5 $\sigma$ confidence level 31 sources at 250 $\mu m$, 31 at 350 $\mu m$ and 25 at 500 $\mu m$.

\section{SED fitting}
\label{sed:fit}

\begin{figure*}[h!t]
\begin{center}
\includegraphics[width=18cm]{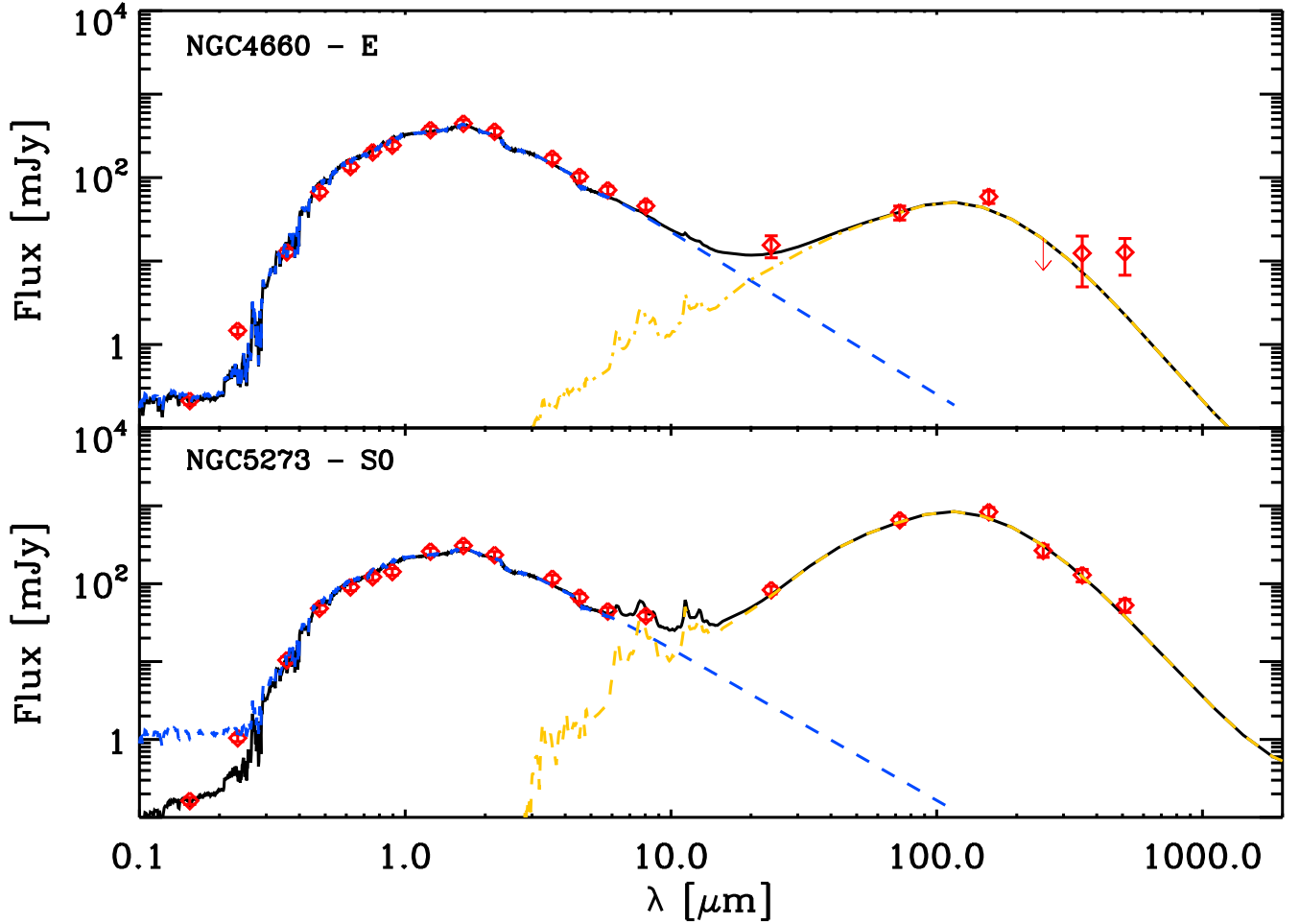}
\caption{Example of fit of SED fit performed with CIGALEMC. For each galaxy, the data are : 2 GALEX, 5 SDSS, 3 2MASS, 4 IRAC, 3 MIPS and 3 SPIRE bands. The top panel shows the fit (black solid line) of NGC4660 (Elliptical galaxy) data (red diamond) with its unabsorbed stellar component (blue dashed line) and the dust emission model (orange dotted dashed line). The bottom panel shows the fit (black solid line) of NGC5273 (lenticular galaxy) data (red diamond) with its unabsorbed stellar component (blue dashed line).}
\label{fig:1}
\end{center}
\end{figure*}

To fit the spectral energy distribution (SED) of our galaxies, we use CIGALEMC\footnote{http://cigale.oamp.fr/} \citep{Serra:11} which is a modified version of the Code Investigating GALaxy Emission (CIGALE \citealt{Noll:09,Giovannoli:11}). CIGALEMC uses a Markov Chain Monte Carlo sampling of the CIGALE parameters which allows to increase the size of the parameter space covered and a more efficient sampling of it.
CIGALEMC uses the \cite{Maraston:05} stellar population model and we use the Salpeter initial mass function (IMF) \citep{Salpeter:55}. The Salpeter IMF is in general a better match for massive early-type galaxies as has been found previously \citep{Grillo:09,Auger:10,Treu:10,Spiniello:11} using stellar dynamics and gravitational lensing. The \cite{Maraston:05} stellar population model includes a realistic treatment of the thermally pulsating asymptotic giant branch (TP-AGB). The TP-AGB phase modelling is important to derive an accurate stellar mass \citep{Maraston:06,Ilbert:10}.
When fitting the data, we assume an exponentially decreasing star formation rate (SFR) for both the old and new star population following \cite{Giovannoli:11}. The age of the old stellar population is fixed to 10 Gyr and the e-folding time of the new population is fixed to 20 Gyr, which corresponds effectively to a constant bursting time \citep{Giovannoli:11,Serra:11}. We use the \cite{Calzetti:94,Calzetti:97} attenuation to describe the dust absorption of star light. We do not add any modification to the Calzetti curve, like a 2175 {\AA} UV bump or a change of slope. The attenuation is modelled independently for the old and young star populations, the attenuation factor for the young population is A$_{ySP}$ (V band attenuation) and there is a reduction factor f$_V$ for the old stellar population (A$_{ySP} \times$ f$_V$).\\
The IR emission from the dust is computed using \cite{Dale:02} model, which is composed of 64 templates parametrized by a slope $\alpha$. This slope represents the power-law slope of the dust mass over the heating intensity. \cite{Dale:02} followed \cite{Desert:90} approach by dividing their dust emission sources into large grains, small grains and PAHs. They normalized these components using observations from IRAS, ISO and SCUBA. CIGALEMC also includes a model for the AGN emission, using the AGN templates from \cite{Siebenmorgen:04a,Siebenmorgen:04b} and a parameter for its amplitude f$_{AGN}$.\\
The eight free parameters of the fit, {$\tau_{old}$, t$_{new}$, f$_{ySP}$, A$_{ySP}$, f$_{V}$ , $\alpha$, f$_{AGN}$, M$_{gal}$, are described in table \ref{tab:1} along with their priors. 

\begin{deluxetable}{lll}[ht!]
\tablecaption{Parameters fitted by CIGALEMC to the galaxy SEDs, \\with priors chosen.}
\tablehead{\colhead{Parameters} & \colhead{Priors} & \colhead{Description}}
\startdata
$\tau_{old}$  &  0 <$\tau_{old}$< 10 Gyr & old star population e-folding time\\ 
t$_{new}$ & 0 <t$_{new}$< 5 Gyr & age of the new stellar population\\
f$_{ySP}$ & 0 <f$_{ySP}$< 1 & fraction of young stars\\
A$_{ySP}$ & 0 <A$_{ySP}$< 5 mag. & dust extinction for young stars\\
f$_{V}$  &  0 <f$_{V}$< 1 & factor for dust extinction of old stars\\
$\alpha$ & 0.06 <$\alpha$< 4 & slope of the dust mass over heating\\
f$_{AGN}$ & 0 <f$_{AGN}$< 1 & AGN fraction of the dust luminosity\\
M$_{gal}$ & 6 <M$_{gal}$< 13 & logarithm of the galaxy mass
\enddata
\label{tab:1}
\end{deluxetable}

In the analysis of the SEDs, we use some derived parameters : SFR, age$_{D4000}$, L$_{bol}$, L$_{dust}$, M$_*$; these are computed either from the fitted parameters and/or from the fitted SED.
The SFR is computed from the contribution of the young and old star populations, however the young star population gives in general the dominant contribution. Therefore the SFR is mostly depending on the fitted parameters M$_{gal}$ the normalization, f$_{ySP}$ the fraction of young star and $t_{new}$ the age of young star population. SFR increases with M$_{gal}$ and f$_{ySP}$, it decreases slightly as young stellar population ages, i.e. as $t_{new}$ gets larger. Apart from its M$_{gal}$ dependency, the SFR is constrained by the UV, optical and NIR part of the spectrum via the stellar population synthesis (SPS) of \cite{Maraston:05}. 
The age$_{D4000}$ is calculated on the unreddened fitted SED by taking the ratio of the average flux per frequency unit of the wavelength ranges 4000-4100 {\AA} (red continuum) and 3850-3950 {\AA} (blue continuum) and matching this ratio to the age calculated on Maraston models \citep{Maraston:05}. L$_{dust}$ is composed of an AGN contribution fitted with a \cite{Siebenmorgen:04a,Siebenmorgen:04b} and a dust component, the dust component part of L$_{dust}$ is constrained by the fitted absorption in UV and optical and the FIR emission consistently. M$_*$ is calculated by integrating over the evolution track of the \cite{Maraston:05} model, and depends primarily on the UV, optical and NIR part of the spectrum, except for the overall normalization defined by M$_{gal}$. L$_{bol}$ is the total luminosity, it is computed by integrating the fitted spectrum and therefore depends on the full SED.\\

Figure \ref{fig:1} shows the SED fit of one elliptical galaxy NGC4660 and one lenticular galaxy NGC5273. The red diamonds represent our data points, the black solid line represents the model spectrum , the blue dashed line represents the unabsorbed stellar spectrum and the orange dotted-dashed line represents the dust emission. Modelled spectra are in pretty good agreement with measurements from 1539 {\AA} GALEX to 500 $\mu m$ Herschel-SPIRE. Dust absorption and emission are significantly larger in the lenticular galaxy NGC5273.\\
Typical constraints provided by CIGALEMC are depicted in figure \ref{fig:2} (NGC4660, corresponding SED is at the top of figure \ref{fig:1}). Diagonal elements of this figure show the marginalized (integrated over all the other parameters) probability distribution function (PDF) of each parameter. The off-diagonal elements show the 2-dimensional marginalized PDF for pairs of parameters and allow to quantify the level of degeneracy between pairs of parameters.
Parameters age$_{D4000}$, L$_{bol}$, M$_*$, SFR, L$_{dust}$  are constrained respectively  at the 11, 14, 16, 53, 87 \% level (median of the relative 68\% confidence interval). Among these parameters, the correlation is fairly small except for the pair (L$_{bol}$, M$_*$) because they depend both mostly on the normalization parameter M$_{gal}$.

\begin{figure*}[h!t]
\hspace{-0.8cm}$\,$\input{fig2}
\caption{Constraints on some of the CIGALEMC parameters (M$_*$,SFR,age$_{D4000}$,L$_{bol}$,L$_{dust}$) for NGC4660. The diagonal (solid black line) plots show the PDFs of each parameter, the rainbow-color plots show the 2-dimension PDFs of one parameter versus another (solid black contours represent the 68 and 95 \% confidence intervals).}
\label{fig:2}
\end{figure*}
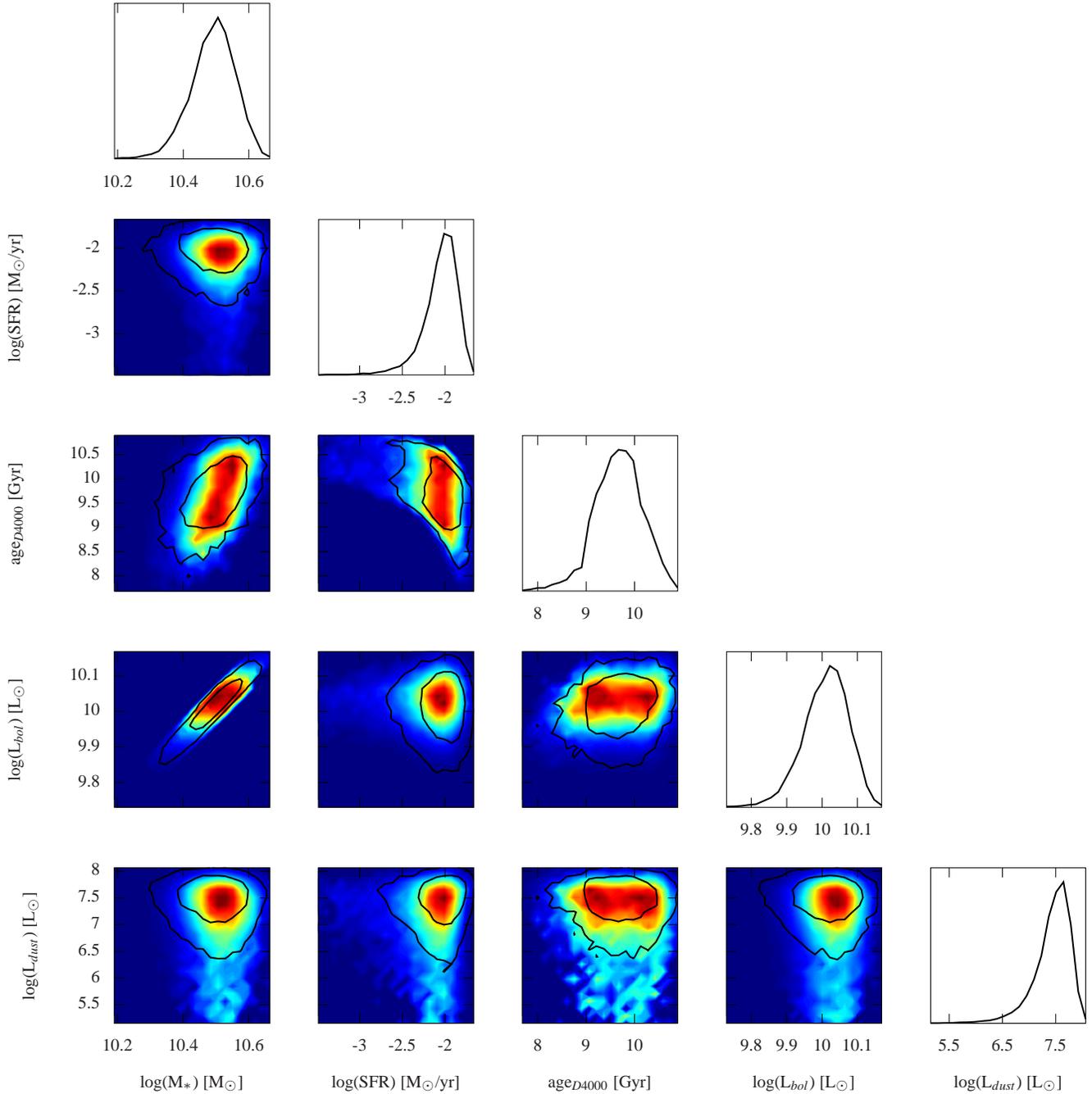

Out of a sample of 221 galaxies (116 E and 105 S0), our model converges to a solution for 193 galaxies. The reduced $\chi^2$ of the fit is between 0.2 and 42, the overall reduced $\chi^2$ is well fitted by a Gaussian distribution centered on 2.5 with a standard deviation of 4.5. We decide to reduce the sample by removing 38 poorly fitted galaxies (reduced $\chi^2$ > 10). Finally, in order to include only strongly constrained galaxy, we select the sub-sample where the galaxy mass is constrained by better than a factor 10 at 68\% c.l., this removes 12 additional galaxies. We check that this last cut does not bias the selection by selecting only massive galaxies. Our reduced sample is composed of 75 elliptical galaxies (type < -3) and 68 lenticular galaxies (type > -3), with a respective average reduced $\chi^2$ of 4.7$\pm$2.6 and 3.9$\pm$2.4, lenticular galaxies are a better fit to our model on average.\\
To check our CIGALEMC estimate of SFR in our galaxy sample, we use the SFR estimate at 24 $\mu$m of \cite{Calzetti:07}:
\begin{equation}
SFR(M_\odot yr^{-1})=1.27 \times 10^{-38}\left[L_{24\mu m} (erg s^{-1})\right]^{0.8850}
\end{equation}
We find a good agreement with the 143 galaxies for which we obtain a SFR as indicated by figure \ref{fig:3}
(left).
\begin{figure*}[h!t]
\begin{center}
\hspace{-0.5cm}$\,$\includegraphics[width=9cm]{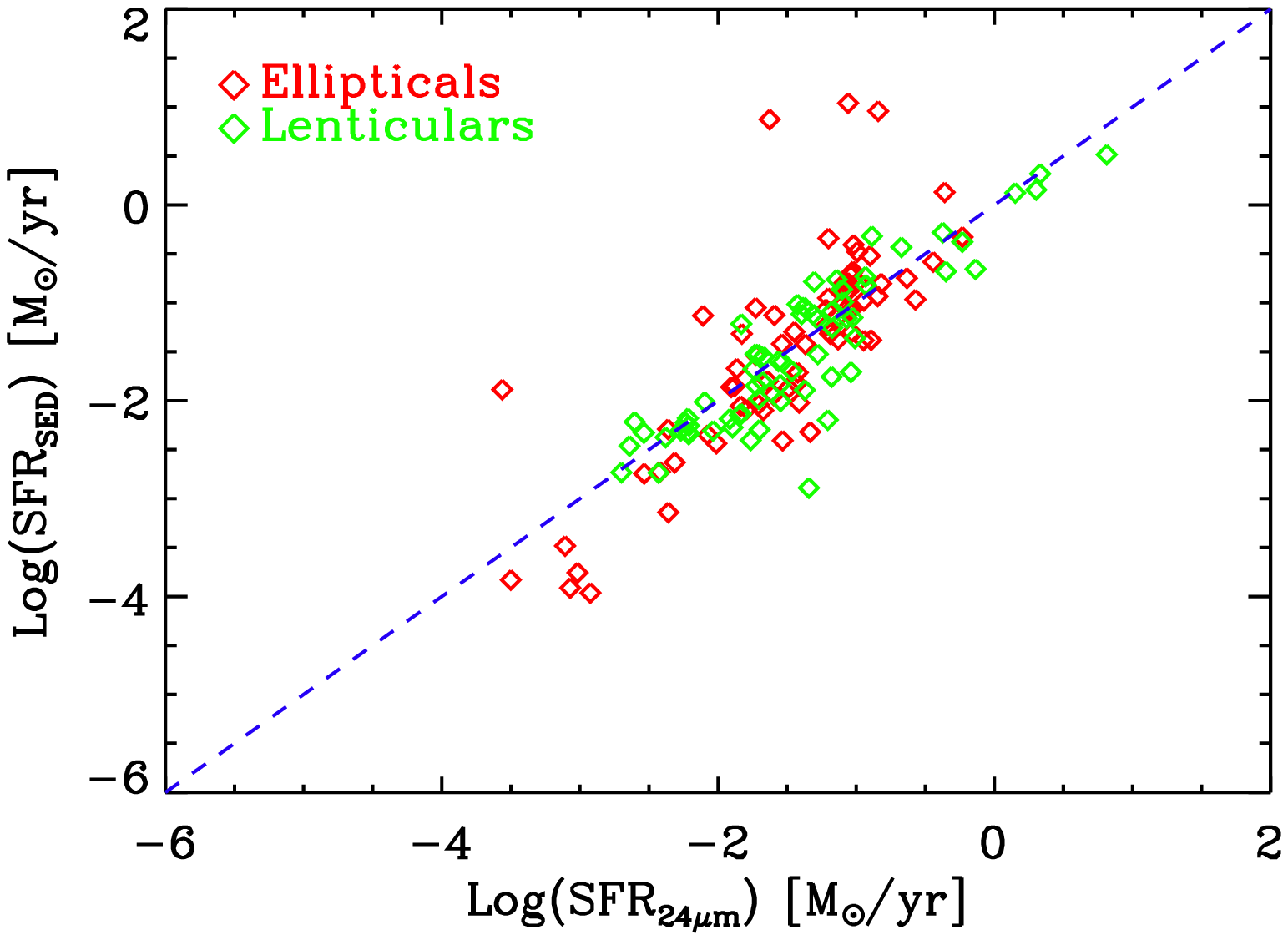}
\includegraphics[width=9cm]{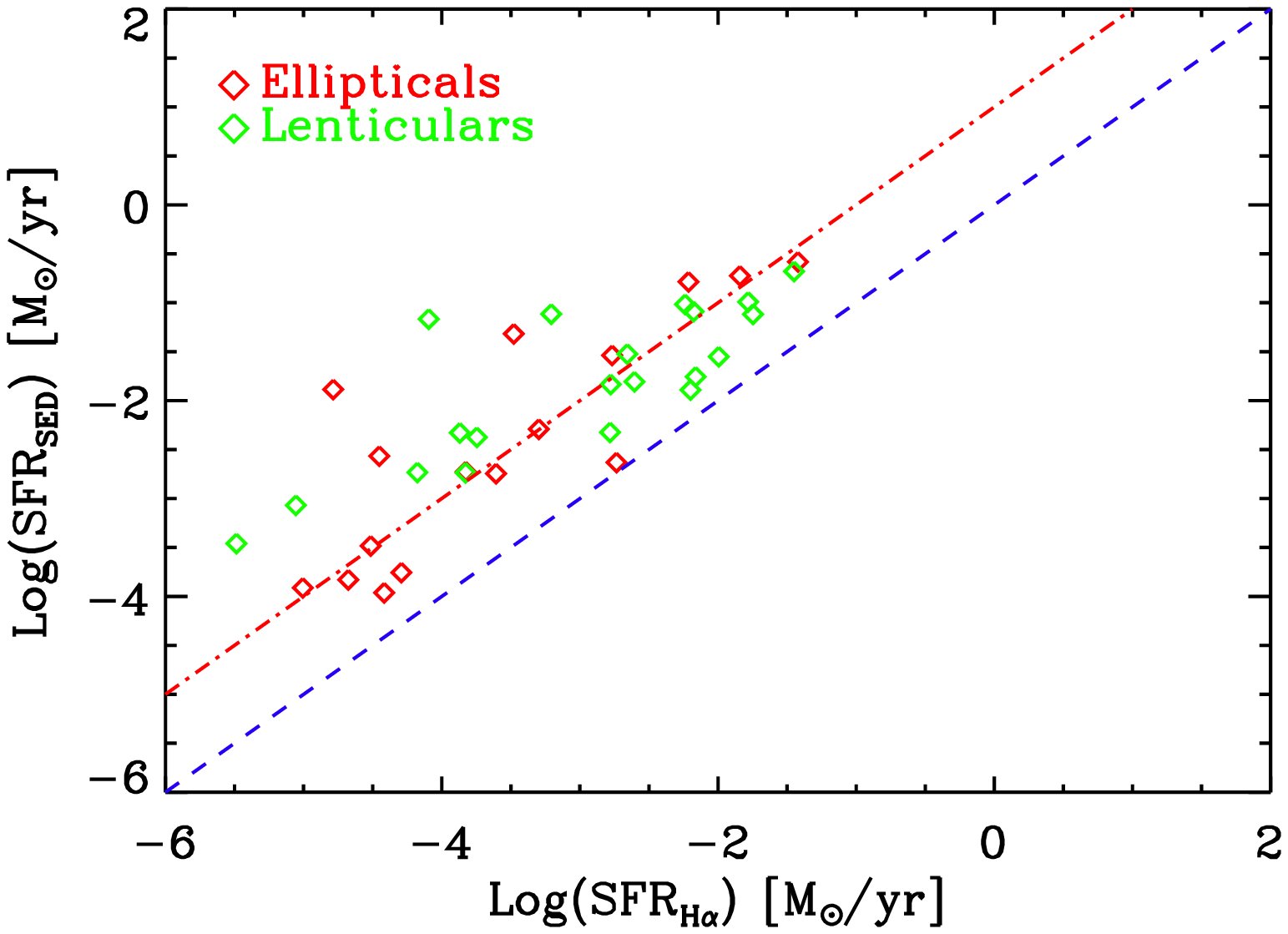}
\caption{{\bf Left :} CIGALEMC estimate of the SFR of our 143 galaxy sample versus the 24 $\mu$m estimate as 
calibrated in \cite{Calzetti:07}. {\bf Right :} CIGALEMC estimate of SFR for a 50 galaxy subset compared to
the SFR calculated from SDSS spectrometric measurements of the H$\alpha$ flux and the \cite{Calzetti:07} relation. The blue dashed line indicates
where the 2 estimates are equal and the red dotted dashed line indicates when CIGALEMC estimate is ten times larger. Our SFR estimate is clearly correlated to the H$\alpha$ luminosity, but the conversion coefficient in the \cite{Calzetti:07} would need to be multiply by 10 for our sample.}
\label{fig:3}
\end{center}
\end{figure*}

Given that CIGALEMC SFR is constrained by the UV/optical/NIR part of the SED, the agreement with the 24 $\mu$m
estimate give us good confidence in CIGALEMC SFR estimates.
Using DR9 SDSS spectrometric measurement of H$\alpha$ obtained for 50 of our galaxies, we use \cite{Calzetti:07} relation between the H$\alpha$ flux and SFR to compare with the CIGALEMC SFR estimate :
\begin{equation}
SFR(M_\odot yr^{-1})=5.3 \times 10^{-42}L(H\alpha)(erg s^{-1})
\end{equation}
Using H$\alpha$ to estimate the SFR, we see a large difference with our estimate (blue dashed line on the right fig.\ref{fig:3}
compared to the data points), the H$\alpha$ SFR estimate is about a factor 10 time smaller than our SED SFR estimate. \cite{Calzetti:07} proposed to 
add a correction with the 24$\mu$m luminosity (5.3 $\times$ 10$^{-42}$ 
0.031 L$_{24\mu m}$) but this correction is larger than the H$\alpha$ contribution in our sample. It is in fact about the
same amplitude as the 24$\mu$m SFR estimate itself (similar to conclusions of \cite{Temi:09a} on H$\beta$ estimate), and  would only reduce the  H$\alpha$ SFR estimate whereas our plot indicates this estimate needs to be increased. The difference could be explained by the small aperture of SDSS spectroscopic data (around 3 arcseconds) and the fact that \cite{Calzetti:07} relation was estimated on starburst whose star-formation is dominated by the central buldge. Our sources might have a more extended star formation distribution, which is captured by our SED and 24 $\mu m$ estimates but underestimated by the SDSS spectroscopic data and the \cite{Calzetti:07} relation.If this difference of aperture was corrected for, the H$\alpha$ SFR estimate might, on the contrary, overestimate the true SFR due to the gas photo-ionization by post-AGB stars \citep{Yan:12,Sarzi:10} and an additional correction would be required. Indeed \cite{Yan:12} found a variation of the emission line ratio as a function of radius in early-type galaxies that could potentially be explained by post-AGB stars \citep{Binette:94} if they are more abundant or more closely distributed with respect to the gas than expected. Previously, \cite{Sarzi:10} also concluded that post-AGB stars are their favorite candidate to power the ionised-gas emission in ETGs based on the correlation of the H$\beta$ line and the stellar surface brightness.
Nevertheless H$\alpha$ luminosities correlate well with our SFR estimates
and changing the relation to SFR(M$_\odot$ yr$^{-1}$)=5.3 $\times$ 10$^{-41}$L(H$\alpha$)(erg s$^{-1}$) leads to a good agreement with our sample (red dotted dashed line on the right of fig. \ref{fig:3}). This implies that the star formation mechanism in the central part of our galaxies and in their outskirt is related and that most likely the star formation is distributed relatively homogeneously.

\begin{figure*}[h!t]
\begin{center}
\hspace{-0.5cm}$\,$\includegraphics[width=9cm]{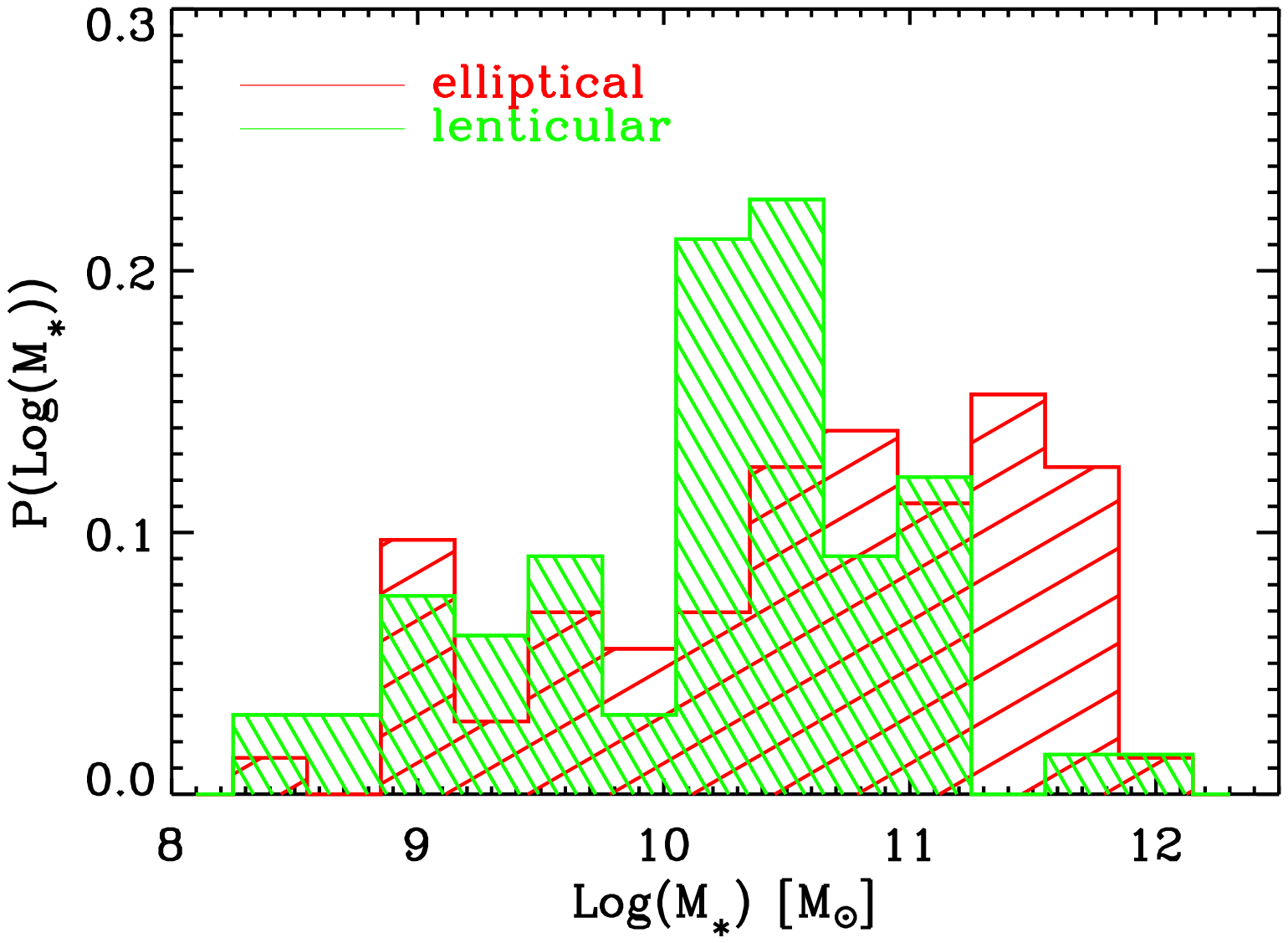}
\includegraphics[width=9cm]{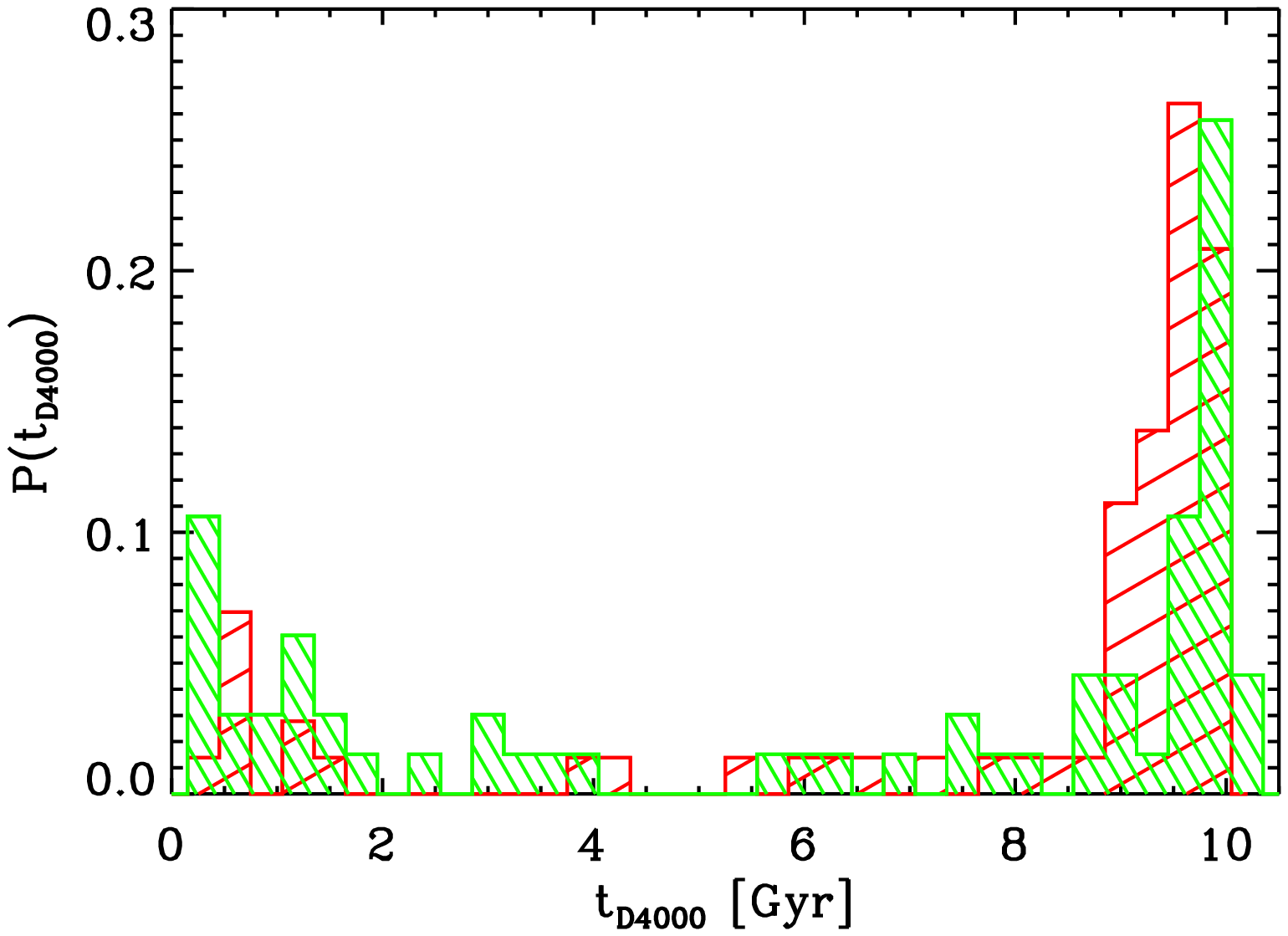}\\
\hspace{-0.5cm}$\,$\includegraphics[width=9cm]{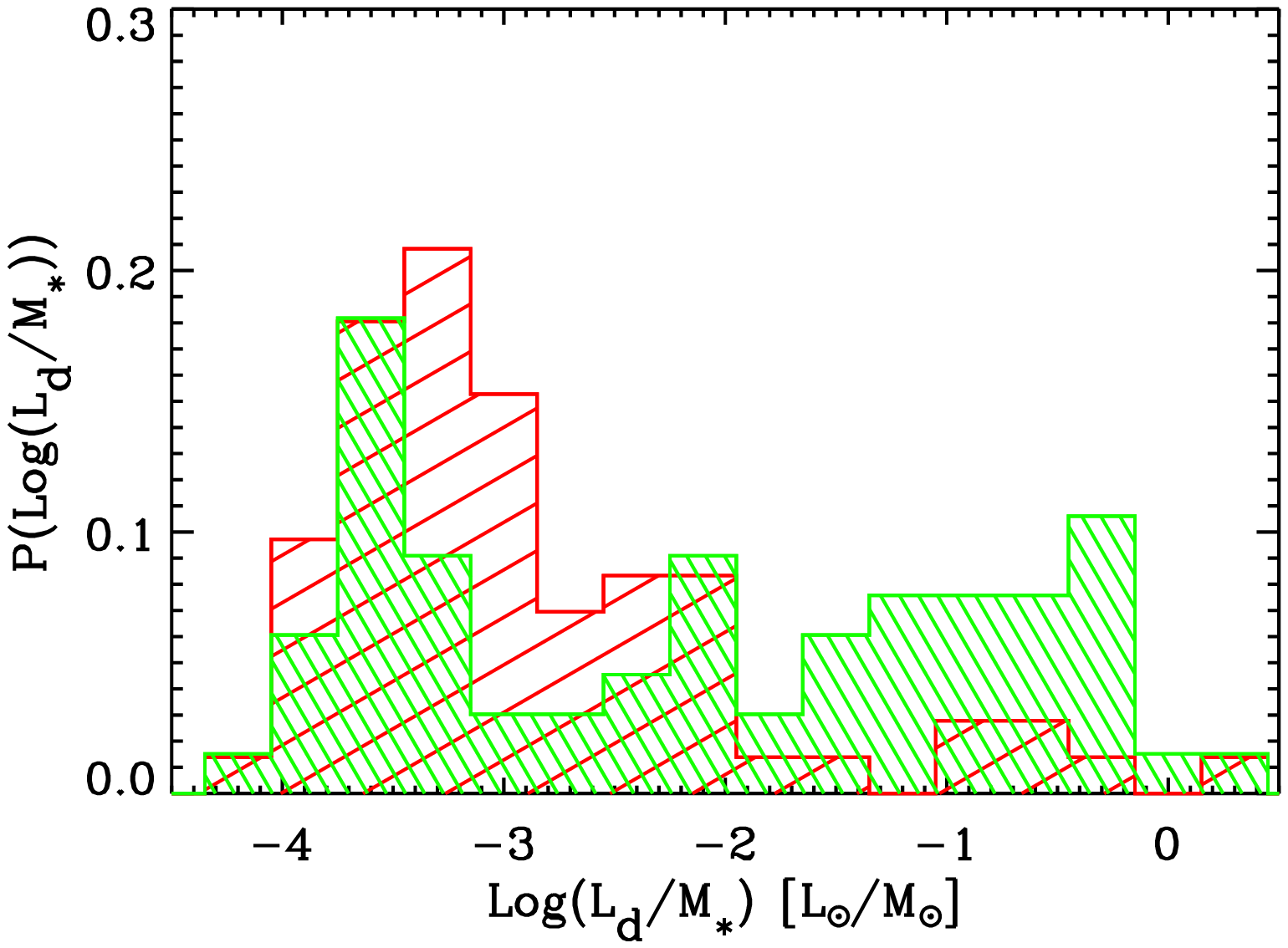}
\includegraphics[width=9cm]{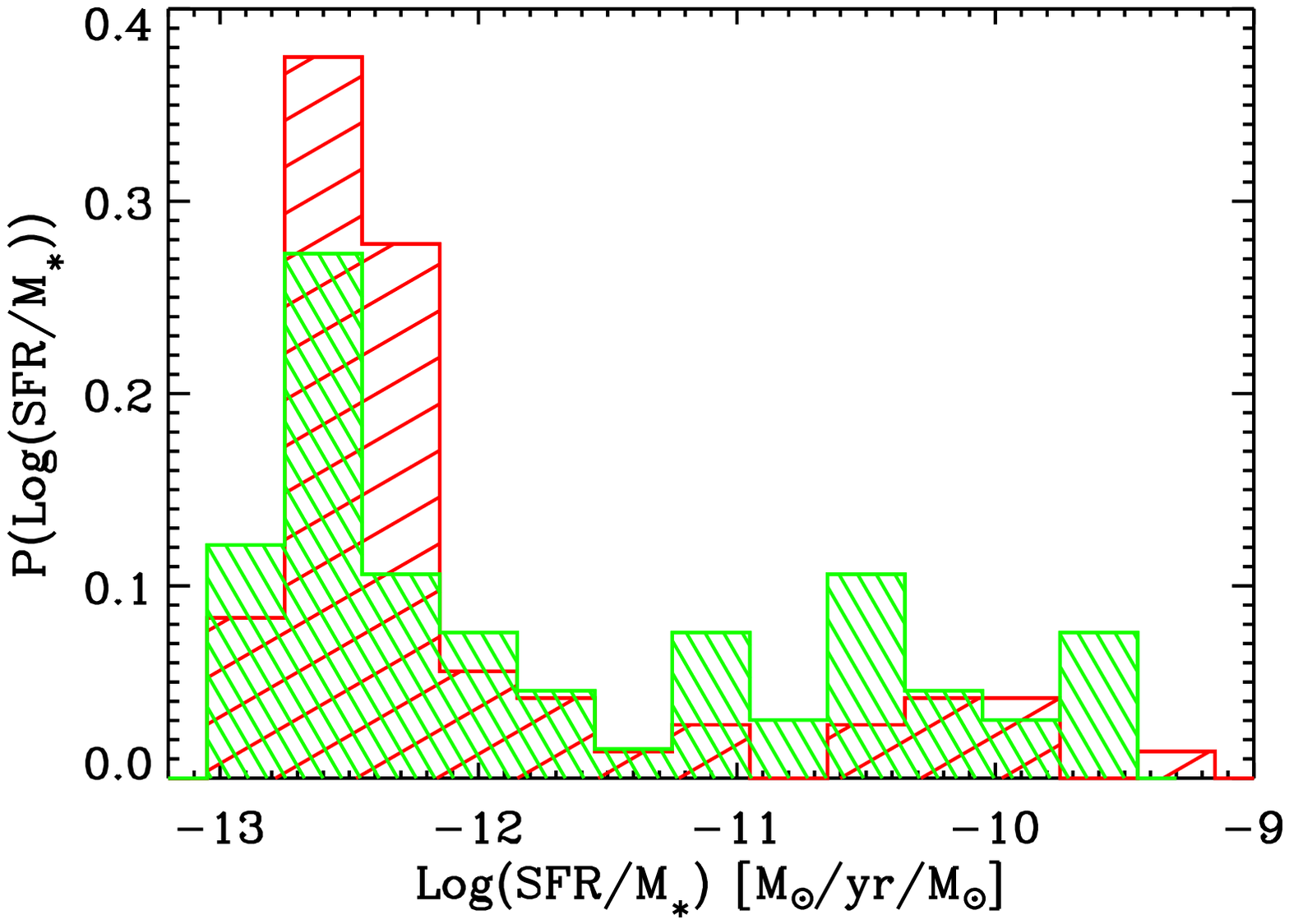}
\caption{Distribution (renormalized to one) of the fitted parameters (from top to bottom and left to right : M$_*$,  t$_{D400}$, L$_d$/M$_*$, SFR/M$_*$) for our galaxy sample}
\label{fig:4}
\end{center}
\end{figure*}

The distribution of the value of our parameters in our reduced sample of galaxies is shown on figure \ref{fig:4}.  Ellipticals have more massive galaxies than lenticulars, about 40\% of ellipticals are heavier than 10$^{11}$ M$_{\odot}$ but only 12\% of lenticulars are. The distribution of age$_{D4000}$ fitted on our galaxies is heavily clustered around 10 Gyr for both type, however there is more young (age$_{D4000}$ < 8 Gyr) lenticulars (46\% of S0) than ellipticals (23\% of E) . When normalizing the SFR and dust luminosity by the stellar mass of each galaxy, the lenticular galaxies are on average brighter and producing more stars. The proportion of E galaxies with a sSFR greater than 10$^{-11.5}$ M$_{\odot}$/yr/M$_{\odot}$ about 16\%, and the proportion of E galaxies with a dust luminosity to stellar mass ratio greater than 10$^{-2}$ is about 15\%. For S0 galaxies, these proportions are respectively about 35\% and 47\%. These results are in agreement with previous studies using FIR photometric indicators \citep{Temi:09a,Temi:09b}, which showed that lenticulars are on average more star-forming and contain more dust. We also compared the sSFR with the slow/fast rotator classification of \cite{Emsellem:11}, which classified galaxies with an angular momentum $\lambda_R$ greater than 0.31$\sqrt{{\epsilon}_e}$ as fast rotators. The number of galaxies, common to $ATLAS^{\rm 3D}$ and our reduced sample of reasonnably fitted galaxies (cf. previous $\chi^2$ and galaxy mass constraints) is 46, with 11 slow rotators and 35 fast rotators. All the slow rotators except for one galaxy are ellipticals, 24 fast rotators are lenticulars. Among the 46 galaxies, only 3 galaxies have a sSFR greater than 10$^{-11.5}$ M$_{\odot}$/yr/M$_{\odot}$. All of these galaxies are fast rotators (3/35), however the small size of this sample does not allow to make a strong conclusion.

\section{Dust Mass and Temperature}
\label{sec:dustmass}
To understand better the dust content of our sample, we decide to fit the far-infrared part of the spectrum with a variety of models parametrized by dust emissivity, temperature and mass. Since CIGALEMC does not provide a direct estimate of the dust mass nor temperature, we apply two additional approaches to measure these quantities : i) we run another SED modeling software MAGPHYS\footnote{http://www.iap.fr/magphys/magphys/MAGPHYS.html} \citep{Dacunha:08}, and ii) we fit some modified black-body spectra to the Far-Infrared portion of our data ($\lambda$ > 60 $\mu$m).

\subsection{Full spectrum fit : MAGPHYS}
Multi-wavelength Analysis of Galaxy Physical Properties (MAGPHYS) is a model able to interpret simultaneously the ultraviolet, optical and IR emission from galaxies and  determine the dust temperature. It separates contribution from dust in stellar birth clouds (BCs) and the ambient interstellar medium (ISM). The dust emission is decomposed in three components for the BCs. One models the PAHs and the NIR emission. The second corresponds to the hot mid-IR continuum emission, described as the sum of two greybodies of temperature of 130\,K and 250\,K respectively. The last component is dust grains in thermal equilibrium, with a temperature allowed to vary between 30 and 60\,K. For the ambient ISM, to keep the number of free parameters manageable, the proportions of these three components are fixed and the temperature of the warm dust in the thermal equilibrium is fixed to 45\,K. \citet{Dacunha:08} added a cold grain component in the ISM with a temperature that can varies from 15 to 25\,K.

Whereas MAGPHYS fits a lot of parameters, our goal is to compute the cold dust temperature and mass of our sample and we concentrate on these two parameters. All the galaxies in our sample are fitted using MAGPHYS, but several parameters are sometimes not well constrained even if the reduced $\chi^2$ of the fit seems acceptable. For instance, among the 221 sources without SPIRE data, 158 sources have good constraint on M$_{dust}$, 132 on L$_{dust}$ and only 36 (16\% of the sample) on T$_C^{ISM}$, hereafter T$_{dust}$. Sources with SPIRE data available return similar results except for the temperature : among 75 sources in the initial SPIRE sample, 45 constrain the M$_{dust}$ parameter, 30 constrain L$_{dust}$  and 20 constrain T$_{dust}$.
FIR-data are important to constrain T$_{dust}$ accurately. 50\% of the sources are constrained when the full FIR spectrum is available and less than 25\% are constrained when SPIRE data are missing. This emphasizes the critical need of FIR data when determining the dust temperature using the MAGPHYS model.

Among the galaxies constrained in temperature, we compare the value of T$_{dust}$ obtained with MAGPHYS with and without SPIRE data. There is a huge discrepancy between these estimates and they do not correlate with each other. This result underlines the fact that values of T$_{dust}$ will not be reliable without observations at FIR wavelengths. We estimate that in our sample, only the temperature of the 20 sources with SPIRE data are accurate enough when using MAGPHYS. This encouraged us to explore other methods to determine the dust temperature of our sample, as detailed in the following sub-section.\\

\subsection{Far-Infrared fit}

\begin{figure}[h!t]
\hspace{-0.5cm}$\,$\includegraphics[width=9cm]{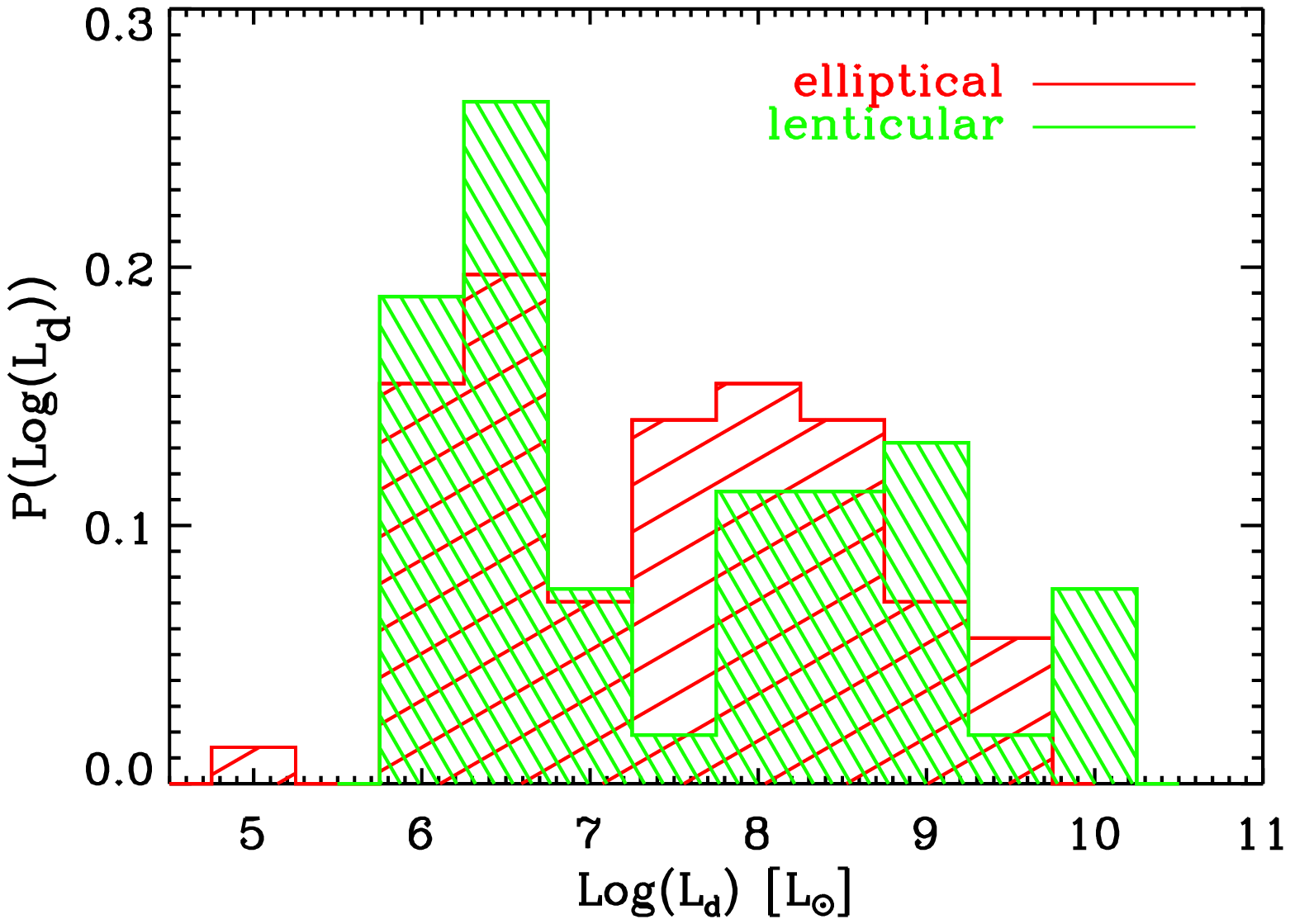}

\hspace{-0.5cm}$\,$\includegraphics[width=9cm]{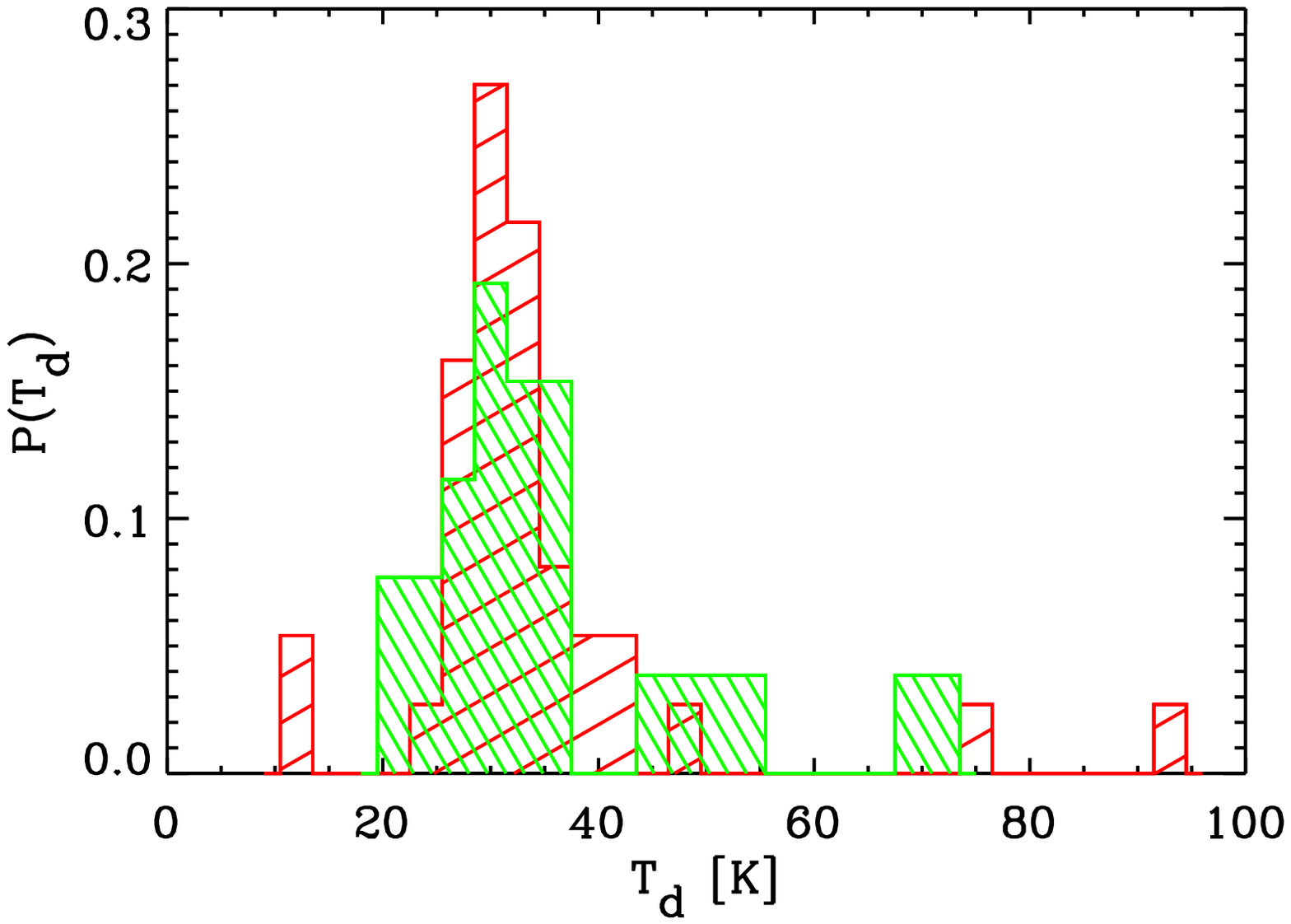}

\hspace{-0.5cm}$\,$\includegraphics[width=9cm]{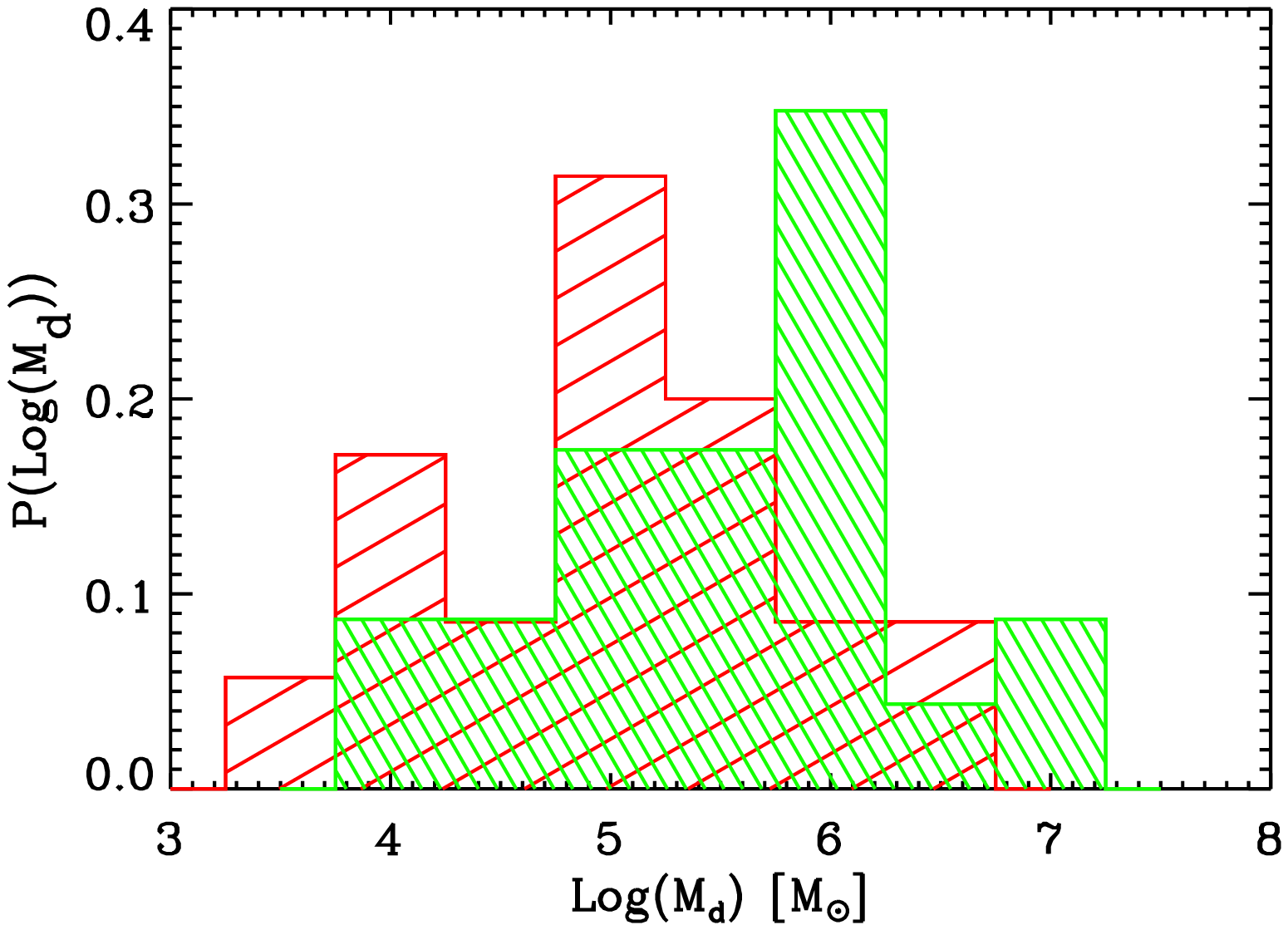}
\caption{Distribution of the dust luminosity (top), temperature (middle) and mass (bottom), obtained
by fitting our one temperature model with $\beta$=1.5 to our data. The elliptical and lenticular
galaxy histograms are respectively in solid black line and dashed blue line, they both are normalized
to one. A few more lenticulars than ellipticals reach a dust mass of $10^6$ M$\odot$.}
\label{fig:5}
\end{figure}

Using only the Far-Infrared part of our data ($\lambda >$ 40$\mu$m), contrarely to MAGPHYS fitting method,
we perform two types of fit on our sources :
\begin{itemize}
\item{a 2 temperature fit with a standard emissivity $\beta$ for big dust grain of 1.5 or 2.0 \citep{Draine:84,Agladze:96,Mennella:98,Reach:95,Boulanger:96,Dunne:01}, on IRAS 60, 100 $\mu m$, MIPS 70, 160 $\mu m$ and 250, 350, 500 $\mu m$ SPIRE data, the luminosity is then :\\
\begin{equation}
L(\nu) \propto \alpha B(\nu,T^1_d) \nu^\beta + (1-\alpha) B(\nu,T^2_d) \nu^\beta
\end{equation}
where B($\nu$,T) represents a blackbody spectrum of temperature T at frequency $\nu$,
$\alpha$ gives the luminosity ratio of the two dust species. T$^1_d$ is constrained between
10 and 45 K, T$^2_d$ between 45 and 95 K.
}
\item{a single temperature fit with emissivity $\beta$, either 1.5 or 2.0, on IRAS 60, 100 $\mu m$, MIPS 70, 160 $\mu m$ and 250, 350, 500 $\mu m$ SPIRE data. T$_d$ is constrained between 10 and 95 K, and the luminosity is then expressed as :\\
\begin{equation}
L(\nu) \propto B(\nu,T_d) \nu^\beta
\end{equation}
}
\end{itemize}

\begin{figure}[h!t]
\hspace{-0.5cm}$\,$\includegraphics[width=9cm]{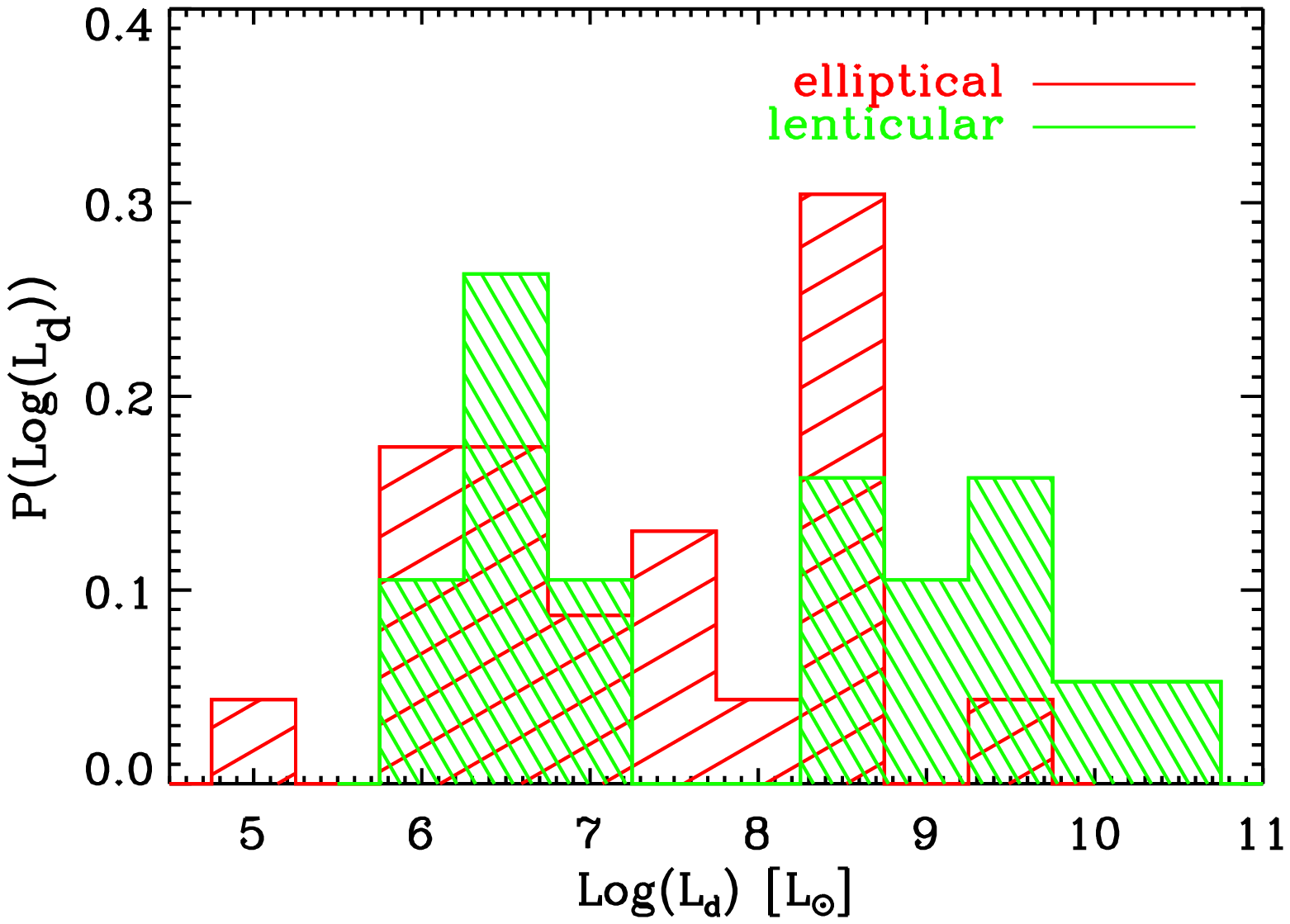}

\hspace{-0.5cm}$\,$\includegraphics[width=9cm]{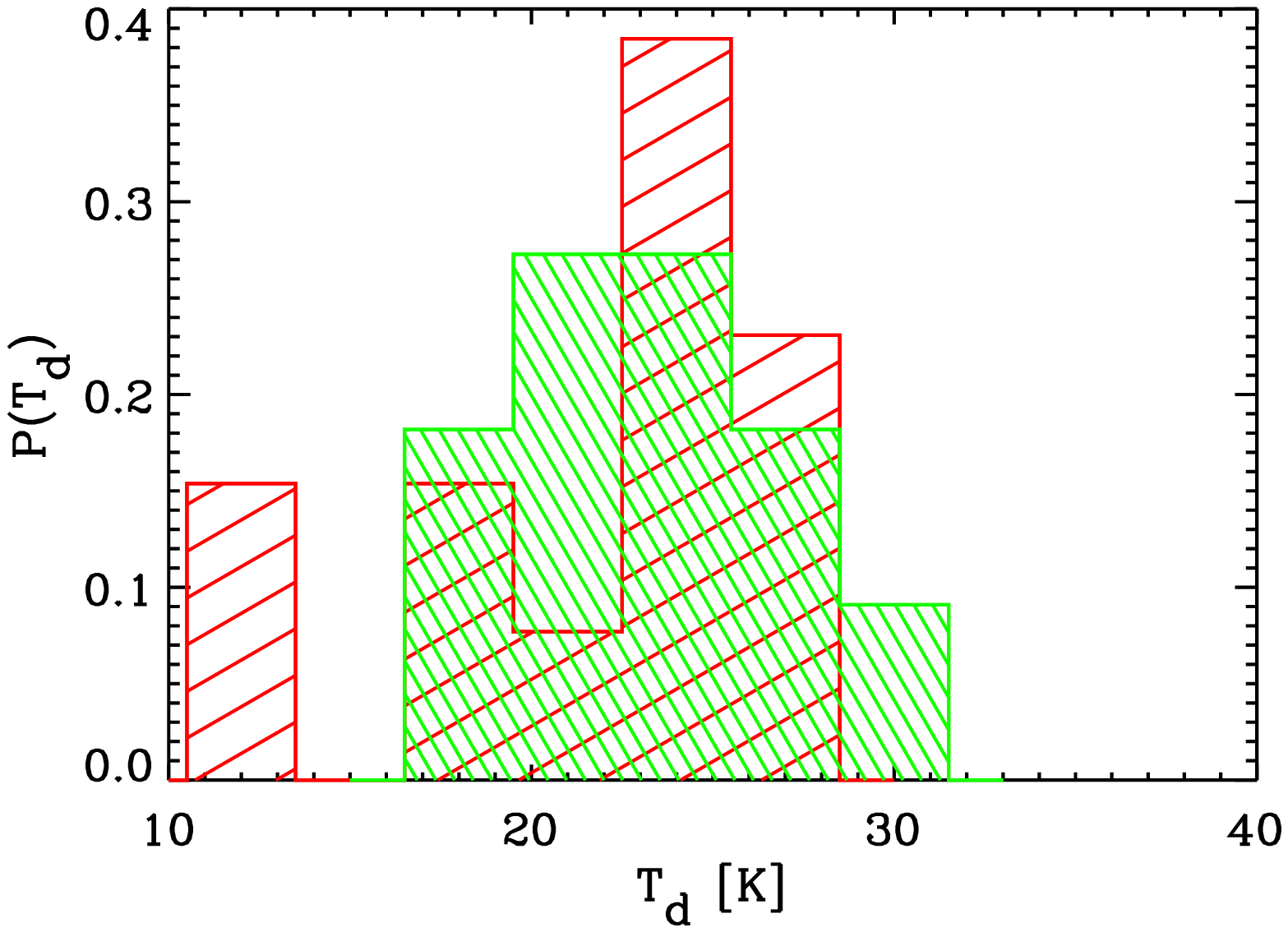}

\hspace{-0.5cm}$\,$\includegraphics[width=9cm]{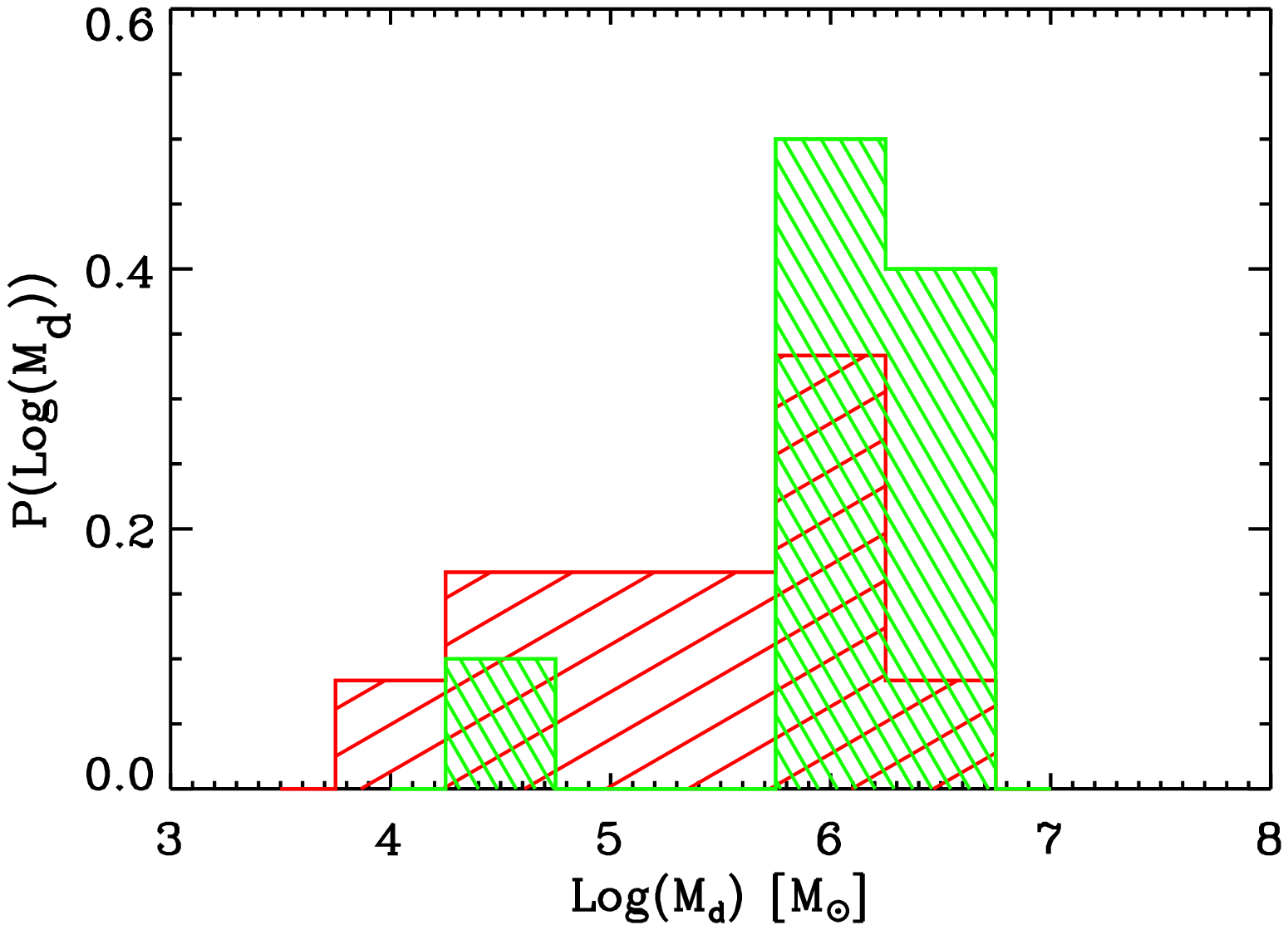}
\caption{Distribution of the dust luminosity (top), cold temperature (center) and mass (bottom), obtained
by fitting our two temperature model with $\beta$=2.0 to our data. The elliptical and lenticular
galaxy histograms are respectively in solid black line and dashed blue line, they both are normalized
to one. Lenticulars appear slightly more dusty and more luminous.}
\label{fig:6}
\end{figure}

We compute fluxes in each band by integrating over the various instrument filters f$_i$, which
have been normalized so that $\int{f_i(\nu) d\nu} = 1$. We normalize fluxes with the
total dust luminosity defined as $L_d = \int{L(\nu)d\nu}$. Fluxes in each band, F$_i$ are then defined as :\\
\begin{equation}
F_i = \frac{L_d}{4\pi D_l^2} {\int{L(\nu) f_i(\nu) d\nu}\over \int{L(\nu)d\nu} }
\end{equation}
To obtain the mass of the dust we assume that dust grains are in thermal equilibrium and use the \cite{Hildebrand:83} formula, which gives a relation between the dust temperature (T$_d$), the dust luminosity (L$_d$) and its mass (M$_d$):
\begin{equation}
L_d(\lambda)=4\pi M_d\kappa(\lambda)B(\lambda,Td)
\end{equation}
We use the same hypothesis as \cite{Dunne:00} and take $\kappa$ at 850 $\mu$m, $\kappa$(850 $\mu$m), to be equal to 0.077 kg$^{-1}$/m$^2$ (based on the calculations of \citealt{Draine:84,Hughes:93}). We assume the emissivity to follow the relation $\kappa(\lambda) = \kappa(850 \mu m) (\lambda/850 \mu m)^{- \beta}$, the dust mass can then be expressed as :
\begin{equation}
M_d=L_d/\int{4\pi\kappa(\lambda) B(\lambda,T) d\lambda}
\end{equation}

The results of the fit can be found in figures \ref{fig:5} \& \ref{fig:6} and tables \ref{tab:td1} \& \ref{tab:td2}. When fitting with a two temperature model, our algorithm manage to converge on 42 galaxies (23 E and 19 S0), when using the single temperature model we constrain 124 galaxies (71 E and 53 S0) with a reduced $\chi^2$ of 10 or better. The average reduced $\chi^2$ of our two temperature fit is 3.4, the reduced $\chi^2$ for our single temperature fit is 3.3. Introducing a more complex model does not improve our fit on average, although our frequency coverage is limited in the Far-IR. Some of our galaxies are known to have a strong synchrotron emission (NGC4267 \& NGC4486 for instance) as indicated in \cite{diSerego:13}, they only represent a small fraction of the sample and most of them have a very large reduced $\chi^2$ and are therefore not included in the following analysis.\\ 
The fit is slightly better on average for an 2.0 emissivity than 1.5 in the two temperature but it is the opposite for the single temperature. Changing the emissivity in the two temperature fit from 1.5 to 2, increases the luminosity by 20\%, decreases the cold temperature by 10\% and increases the dust mass by a factor 3. In our single temperature case, we compare the fit with an emissivity $\beta$ of 1.5 and 2 and find that the temperature and mass change respectively by $\sim$ -10, -40 \% when using 2 instead of 1.5, but the luminosity is barely changed. In the following, we will only discuss our fit with an emissivity 2.0 for the 2 temperature model and 1.5 for the single temperature model.\\
 The distributions in figures \ref{fig:5} \& \ref{fig:6} show that galaxies
 fitted by the 2 temperature model have a lower temperature than the one fitted 
with the single temperature model (22.4 K versus 32.3 K). This can be explained by the 
fact that our 2 temperature model containing four free parameters, it requires 5 data points.
 This means that at least one data point from Herschel-SPIRE is required to fit the 2 temperature model,
whereas the one temperature fit does not require Herschel data with only two free parameters.
The 2 temperature model is only possible on galaxies with colder dust. This result is in 
agreement with \citet{Skibba:11} where they find higher dust to stellar flux ratio for 
Herschel-detected galaxies, because Herschel is tracing additional cold dust that was not
 detected with Spitzer. The 32.3 K average temperature derived on the single temperature
model is very close to average temperatures derived from other galaxy selections, like for
instance the 35 K estimate from \cite{Elbaz:10} and the 28 K estimate from \cite{Amblard:10}.
Figure \ref{fig:5} \& \ref{fig:6} reveal also that lenticular
 galaxies have slightly more dust than elliptical (more apparent in 2 
temperature model), and slightly higher dust luminosity when their SED is modelled 
with 2 temperatures.

Fig.\,\ref{fig:7} shows the evolution of the ratio of dust mass 
to stellar mass in logarithmic scale versus the stellar mass in elliptical galaxies (in red) and
 in lenticular ones (in green) with the single temperature model, when selecting only sources with 
a secure dust mass (less than 50\% relative error) and a good fit to the full SED and the FIR 
spectra (34 sources). We decide to present results only for the single temperature model,
 we have similar results but with poorer statistics for the 2 temperature model.\\
 The dashed lines represent the best linear fit obtained  with these 34 sources (black), with 
elliptical galaxies among this sample (red) and with lenticular galaxies (green). 
The general trend is that the ratio M$_{dust}$ 
to M$_{star}$ decreases with increasing stellar mass. This relation is stastically identical 
for both elliptical  galaxies (red dashed line) and lenticular ones (green dashed line),
the slopes being -1.16\,$\pm$0.16, -0.97\,$\pm$0.18, -1.20\,$\pm$0.30 for the whole sample, the elliptical
 galaxies and for the lenticular galaxies respectively. The decreasing trend is in good agreement with other studies, measurements from \cite{Smith:12} seem compatible with our fit although their stellar mass range is very small (10$^{10}$ to 10$^{11}$ M$_\odot$).\cite{Agius:13} found a shallower slope of -0.55 with larger dust-to-stellar mass ratio (10$^{-2.5}$ to 10$^{-4}$), it uses a sample detected by the shallow H-ATLAS survey which could explain that it selects higher dust-to-stellar mass ratio. \citep{Rowlands:12} using data from H-ATLAS as well found higher dust masses, around 7$<$log(M$_{d}$/M$_{\odot}$)$<$9 and higher dust-to-stellar mass ratio. \citet{Skibba:11} measured  dust-to-stellar mass ratio in the same range as ours. \cite{Cortese:12} measured dust-to-stellar mass ratio on late-type galaxies ranging from 10$^{-2}$ to 10$^{-4}$ and on early-type galaxies from 10$^{-3.5}$ to 10$^{-6}$.

\begin{figure}[ht!]
\hspace{-0.5cm}$\,$\includegraphics[width=9cm]{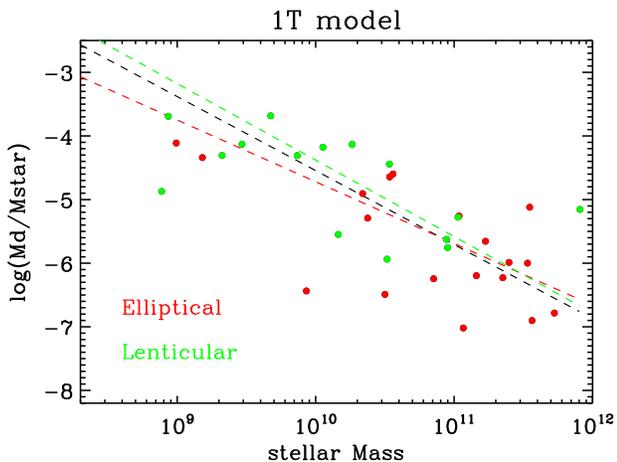}
\caption{Ratio of dust mass to stellar mass in logarithmic scale versus the stellar mass for Elliptical galaxies (in red) and for lenticular ones (in green) for sources with a secure fit (dust mass estimate with a relative error of 50\% or less) with the 1T model. The dashed lines represent the best linear fit obtained  for the whole sample (black), for elliptical galaxies (red) and for lenticular galaxies (green).}
\label{fig:7}
\end{figure}

Fig.\,\ref{fig:8} shows the evolution of the specific star formation rate as a function of the 
dust-to-stellar mass ratio for Ellipticals (red triangles) and for Lenticulars (green triangles)
in logarithmic scale. Fig.\,\ref{fig:8} is similar in shape to the banana-shaped plots of \cite{Temi:09b} which show the relashionship between log(L$_{24}$/L$_K$) and log(L$_{FIR}$/L$_K$), and the information it contains is correlated with our plot. However Fig.\,\ref{fig:8} permits a better distinction of SFR and dust mass than the 24 $\mu$m and FIR (70 or 160 $\mu$m) luminosities could.\\ 
The sSFR is constant (log(sSFR)$\sim$ -12.5 M$_{\odot}$/yr/M$_{\odot}$) 
for a dust-to-star mass ratio lesser than 10$^{-5}$. This part of the diagram is mainly populated by 
Elliptical galaxies (68\% of sources). For a higher dust-to-star mass ratio (log(M$_{d}$/M$_{*}$)>-5), 
which mainly contains S0 galaxies (72\% of sources), the sSFR rise steeply with the dust to star mass ratio.
A single outlier source, NGC4494, lies at Log(sSFR) $\sim$ -10 and Log(M$_d$/M$_*$) $\sim$ -6.
This Elliptical galaxy has a high specific star formation rate but a surprisingly small dust-to-star 
mass ratio. NGC4494 contains an AGN according to the Veron catalog \cite{Veron:10} and could
have undergone a recent merger\citep{OSullivan:04}, which might be able to explain the low dust mass.
However, we do not have GALEX or SDSS data for this galaxy, so the fit might have converged to
an unphysical solution.

\begin{figure}[ht!]
\hspace{-0.5cm}$\,$\includegraphics[width=9cm]{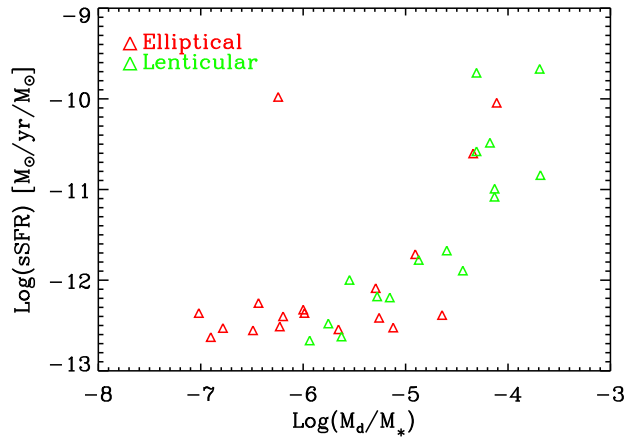}
\caption{Specific star Formation estimated with our CIGALEMC fit versus the dust mass to stellar mass ratio,
estimated with the 1 temperature FIR model. We select data with a reduced $\chi^2$ lower than 10,
a dust mass error of less than 50\% and a dust luminosity logarithm error of less than 1 dex. Elliptical
galaxies are indicated with red triangles, Lenticular galaxies with green triangles. The isolate triangle
at Log(sSFR) $\sim$ -10 and Log(M$_d$/M$_*$) $\sim$ -6 represents NGC4494.}
\label{fig:8}
\end{figure}

 We compared our E/SO classification with the slow/fast rotator one from \cite{Emsellem:11} and obtained the same sample of 46 galaxies of section \ref{sed:fit}, that are in $ATLAS^{\rm 3D}$ and well fitted in our sample. Out of 35 fast rotators, 6 have a  Log(M$_d$/M$_*$) greater than -5.5 but no slow rotator has such a high dust mass. This would indicate that fast rotators are the only one with a large amount of dust, although the small size of this sample does not allow to conclude.

\section{AGN activity}
\label{sec:agnactivity}

AGN activity can influence the star formation rate of galaxy. In order to better understand the variability in our sample of ETGs, we try to detect galaxies in our sample with a significant trace of AGN activity.

\subsection{Spectroscopic selection}

To select AGN, we download the DR9 SDSS spectroscopic line catalog\footnote{http://www.sdss3.org/dr9/spectro\_access.php}
along with the spectroscopic information file and cross-identify it with our 221 galaxy sample.
We identify 54 galaxies (27 S0 and 27 E, compared to the 147 galaxies for which we have photometric data), from which we select the ones with at least 4 lines with a signal-to-noise
ratio of at least 3 out of the 7 lines that we use to produce the Baldwin, Phillips, Terlevich (BPT) diagrams \citep{Baldwin:81}, namely OIII
(5007 \AA), NII(6584 \AA), SII(6718 and 6731 \AA), H$\alpha$ (6563 \AA), H$\beta$ (4861 \AA), OI (6300 \AA). 
We obtain 21 galaxies (13 S0 and 8 E, $\sim$ 40\% of the sample) with enough optical lines detected,
about the same rate as \cite{Rowlands:12} (45\%) and twice the rate of \cite{Schawinski:07b} ($\sim$ 20\%).
Our 24$\mu$m selection and the 250$\mu$m selection of \cite{Rowlands:12} are more likely to select active
ETGs than the optical selection of \cite{Schawinski:07b}.\\
Using the 3 BTP diagrams of figure \ref{fig:9}, with Star-Forming (SF), AGN and LINER area defined by \cite{Kewley:01,Kewley:06}, we identify 5 AGNs (5 S0), 11 SFGs (7 S0 and 5 E) and 4 (1 S0 and 3 E) undetermined galaxies 
(with some lines they are identified as AGN and some others as SFG). We obtain a larger proportion
(69\%) of star forming galaxies than the 57\% of the 250$\mu\,$m selected sample of \cite{Rowlands:12}, 
although the difference is not statistically significant since both samples are quite small. \cite{Schawinski:07b}
, using a much larger sample of optically selected ETGs, found 61\% of SFGs among sources with
optical line detection. Starting with an equal number of E and S0 galaxies in the cross-sample with
SDSS spectroscopic data, more S0 galaxies (65\% of the detected galaxies)  have detected emission lines 
than E galaxies and all AGN identified with the BPT diagrams are S0. Both \cite{Rowlands:12} and \cite{Schawinski:07b} did not separate their numbers between S0 and E morphology, but \cite{Schawinski:07b} found a time sequence
for small to intermediate mass ETGs, going from a star-forming phase to an AGN phase to finish into a quiescent state.
Our results would be consistent with such a picture if the S0 dominate the intermediate AGN phase.
Overall, our S0 galaxies with SDSS spectroscopic are 52\% quiescent, 19\% AGN-dominated and 26\% SF-dominated,
our E galaxies are 70\% quiescent, 0\% AGN-dominated and 18\% SF-dominated.
Separating LINERs from other AGNs would not change our results, since our BPT diagrams do not seem to indicate any LINERs in our sample. Two galaxies are potentially borderline LINERs in one BPT diagram but are clearly not LINERs in another one and neither are classified as such in the NED and Hyperleda databases.\\
We mark, on our BPT diagrams, galaxies for which the 24$\mu$m morphology is point-source like 
(red squares). Among galaxies identified with a 24$\mu$m point-source morphology and for 
which we have SDSS spectroscopic data, all of them have detected lines. 
The 24$\mu$m morphology seems to be a good indicator
of galactic activity. They seem to split evenly between AGN emission (2) and SF emission (3),
24 $\mu$m morphology gives a $\sim$50\% accuracy in AGN detection in our sample.

\begin{figure}[h!t]
\hspace{-0.5cm}$\,$\includegraphics[width=9cm]{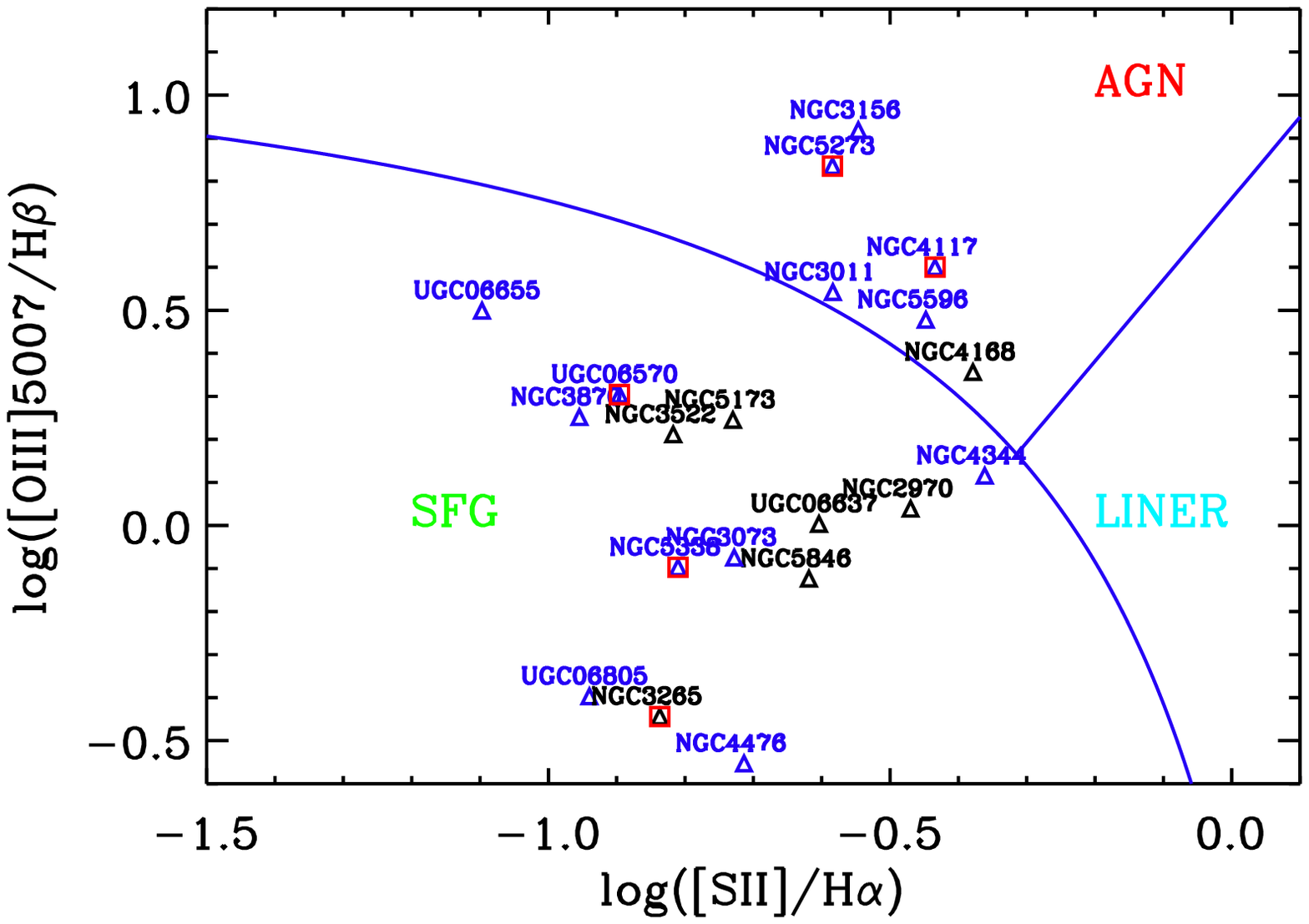}

\hspace{-0.5cm}$\,$\includegraphics[width=9cm]{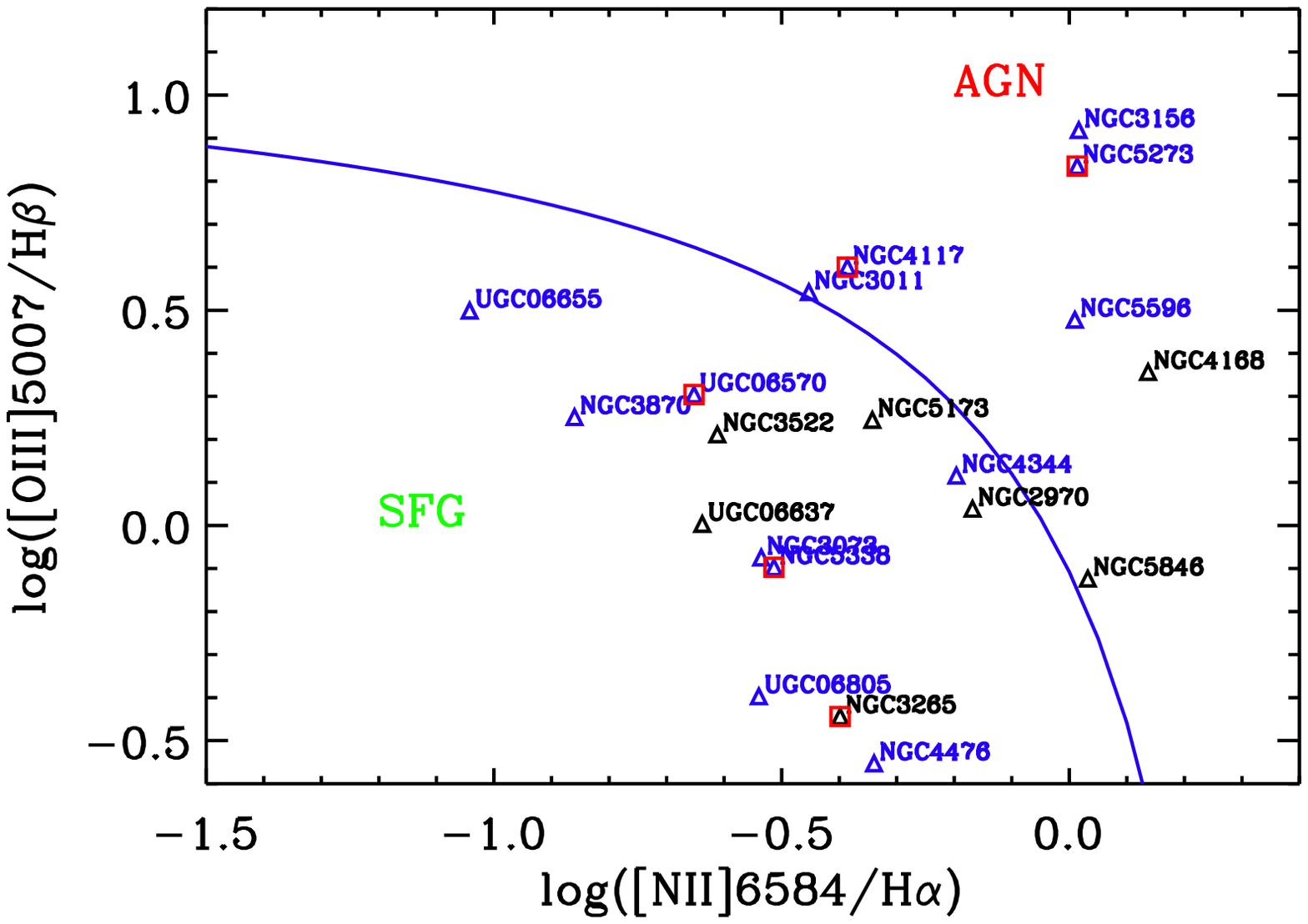}

\hspace{-0.5cm}$\,$\includegraphics[width=9cm]{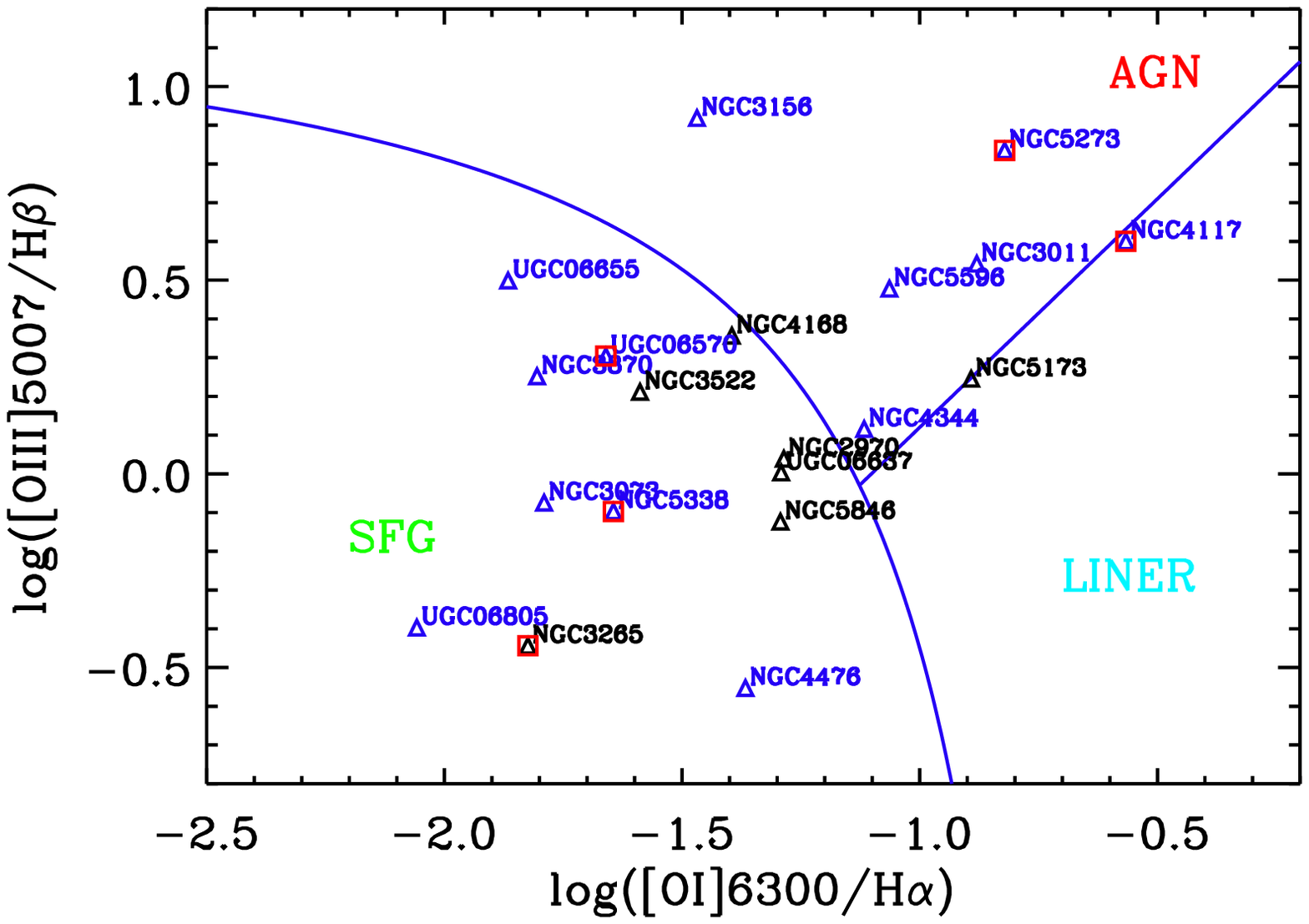}
\caption{BPT diagrams of galaxies in our sample for which we have SDSS spectroscopic data 
and at least for emission lines detected. Lenticular galaxies are represented by blue triangles,
 Elliptical galaxies by black triangles. The blue solid line separates the Star Forming Galaxy (SFG)
region from the AGN one. The red square indicates galaxies which 24$\mu$m morphology is point-like.}
\label{fig:9}
\end{figure}

\subsection{NIR selection}

Another powerful way to select AGN is to use NIR selection. Indeed, the UV to MIR continuum for a galaxy is dominated by a black body emission that would dominate at 1.6$\mu m$, while the UV to MIR continuum of an AGN is dominated by a power law. Different criteria have been proposed 
using  IRAC-IRAC colors \citep[e.g.,][]{Lacy:04, Stern:05} and more recently WISE colors \citep[e.g.,][]{Stern:12}. We apply these three criteria to our sample: 9 sources are selected following \citet{Lacy:04}, only one source (plus a boarder-line source) with \citet{Stern:05} criteria, and 4 sources following \citet{Stern:12} using WISE filters. Very few sources are selected by these criteria, this might be explained by these criteria being designed in order to select highly obscured quasars and our sample of ETGs is quite poor in dust.

\citet{Assef:12} underline the incompleteness and the unreliability of these criteria, stating a reliability of only 36\% for the selection of \citet{Lacy:04} and a completeness of 79\%, a 67\% reliability/ 51\% completeness for \citet{Stern:05} and a 45\% reliability/ 70\% completeness for \citet{Stern:12}. This leads us to the conclusion that we should combine the 3 criteria to increase the reliability of the NIR selection on our sample. We finally have only 2 sources selected as AGN via the NIR-methods: NGC1377 and IC5063. We do not have any SDDS spectra for these two sources so we cannot correlate this NIR classification
with the BPT spectral classification. Both of these sources have a high f$_{agn}$ fraction from the SED-fitting which could confirm their AGN activity.  IC5063 is also detected in X-ray and has a luminosity L$_{X} >$ 10$^{41}$ erg.s$^{-1}$, pointing to some AGN activity. NGC1377 is a power-law source, i.e. presenting Fl$_{IRAC4}>$Fl$_{IRAC3}>$Fl$_{IRAC2}>$Fl$_{IRAC1}$, once again characterizing an AGN. 

Fig.\,\ref{fig:10} shows the distribution of our sample in the \citet{Lacy:04} diagram. Elliptical galaxies are represented with dots while lenticulars are represented with crosses. We add to this figure the information collected on the BPT diagrams with a color code. Black symbols represent galaxies for which we do not have emission lines or SDSS data, blue symbols represent AGN sources, pink ones represents the potential AGN and green ones represent the star-forming sources. Star-forming sources (green dots and crosses) are mainly distributed in the upper part of the diagram on the outside of the \citet{Lacy:04} box. Sources in this part of the color-color plot, with bluer S$_{5.8}$/S$_{3.6}$ colors and redder S$_{8.0}$/S$_{4.5}$ colors than the rest of the sample, are known \citep{Lacy:04} to be low-redshift galaxies with their 6.2 and 7.7 $\mu m$ PAH bands redshifted into the IRAC 8$\mu m$ filter. The presence of PAHs in these sources shows that they are forming stars and  most-likely present emission lines in their spectra.
We also notice that AGNs from the BPT diagram (blue dots and crosses) are outside the NIR-selection, showing that these two methods select different types of AGN with different properties. The NIR-selection will select highly obscured quasars.
Another trend noticeable on figure \ref{fig:10} is the separation between Ellipticals (dots), concentrated in the left-bottom part, and lenticulars (crosses). This shows that lenticular galaxies have more dust than ellipticals, which goes in the same direction that we have already noticed on Figs.\ref{fig:5} \& \ref{fig:6}.

\begin{figure}[ht!]
\hspace{-0.5cm}$\,$\includegraphics[width=9cm]{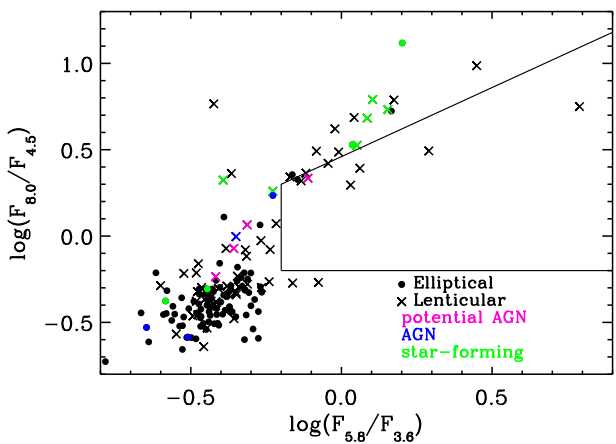}
\caption{Distribution of the sources from our sample in the \citet{Lacy:04} diagram displaying log(S$_{8.0}$/S$_{4.5}$) versus log(S$_{5.8}$/S$_{3.6}$). Elliptical and Lenticulars are represented by dots and crosses respectively. We also show here the AGN activity versus star-forming activity of our sample from our analysis of the BPT diagram: star forming (green), AGN (blue) and potential AGN (pink). The black quadrilateral shape correspond to the \citet{Lacy:04} box where the highly obscured quasars lie.}
\label{fig:10}
\end{figure}

Similarly to the BPT diagram, the NIR selected AGNs in our sample are mainly 
composed of S0 ($\sim$ 90\%). Using this criteria, we find that 12.3\% of S0
are AGN-dominated while only 1.1\% of Elliptical galaxies are AGN-dominated.
As we recalled previously, this estimator is not complete and it could also
be biased for our selection, but it qualitatively matches the proportions
we find with the BPT diagrams (19\% for S0 and 0\% for E).\\ 
As a comparison in table \ref{tab:tab_agn}, we indicate sources detected in X-ray with Chandra 
with a high X-ray luminosity , i.e. L$_X > 10^{41}$ erg.s$^{-1}$, believed to be AGN sources as well. We also show if the SED-fit parameter $f_{agn}$ indicates the presence of an AGN. $f_{agn}$ is not strongly constrained by our SEDs, and we chose to report its value by indicating whether its lower 95\% confidence level limit is greater than 10\% at the 95\% confidence level, ie. the AGN luminosity in MIR contributes at least 10\% of the total IR luminosity. Galaxies with lower AGN contribution in their SED are reported with a ``inc.'' for inconclusive, since we could not conclude for these galaxies how large the potential AGN contribution could be.

\begin{deluxetable}{llllllll}[ht!]
\tablecaption{AGN classification from the 4 criteria : from the BPT diagram, from the AGN template in the SED fit, from the NIR color criteria, from X-ray luminosity. A dash line indicates that we did not have data and ``inc.'' indicates data were inconclusive.}
\tablehead{\colhead{Name} & \colhead{Type} & \colhead{BPT} & \colhead{SED} & \colhead{NIR} & \colhead{Xray}}
\startdata
Eso103-35 &  S0 & -- & -- & yes &  --  \\
Eso428-14 &  S0 & -- & -- & no  & yes  \\
IC4296        &   E  & -- & inc. & no  & yes  \\
IC5063       &  S0 & -- & yes & yes & yes \\    
NGC0221  &   E  & -- & -- & no  & yes  \\
NGC0315  &   E  & -- & inc. & no  & yes  \\
NGC0410  &   E  & -- & inc. & no  & yes  \\
NGC0507  &   E  & -- & inc. & no  & --   \\
NGC0533  &   E  & -- & inc. & --  & --   \\
NGC0596   &  E   & -- & yes & no &  --  \\
NGC0807  &   E  & -- & inc. & --  & --   \\
NGC1016  &   E  & -- & inc. & no  & --   \\
NGC1374  &  E   & -- & inc. & yes &  --  \\
NGC1377  &  S0 & -- & yes & yes & -- \\ 
NGC1386  &  S0 & -- &  inc.  & yes &  no  \\
NGC1407  &  S0 & -- &  --  &  -- &  yes  \\
NGC1439   &  E   & -- & yes & no &   --   \\
NGC1510   &  S0 & -- & inc.  & yes &  --   \\
NGC2832   &  E   & -- & inc.  &  no  &  yes  \\
NGC3011  &  S0 &  yes &  inc.  & yes &  --  \\
NGC3156  &  S0 &  yes &  inc.  & -- &  no  \\
NGC3516  &  S0 &  -- &  inc.  & yes &  --   \\
NGC3593  &  S0  & -- & inc. & no  & --    \\
NGC3610   &  E   & -- & yes &  no &  no   \\
NGC3706   &  E   & -- & yes &  no &  --  \\
NGC4117  &  S0 &  yes &  inc.  & no &  no  \\
NGC4138  &  S0 &  -- &  inc.  & yes &  --  \\
NGC4168  &  E   &  yes &  inc.  &  no &  --  \\
NGC4261  &  E   &  --  &  inc.  &  no &  yes \\
NGC4344  &  S0 & inc.  &  -- & no &  -- \\
NGC4382  &  S0 &  --  & yes &  no  &  no   \\
NGC4460  &  S0 &  --  & inc. &  yes  &  --  \\
NGC4638  &  S0 &  --  & yes &  no  &  no   \\
NGC4915   &  E   &  --  & yes &  no  &  no  \\
NGC5061   &  E   &  --  & yes &  no  &  --   \\
NGC5173  &  E   & inc.  &  inc.  &  no  &  --   \\
NGC5273  &  S0 &  yes &  inc.  &  no  &  yes  \\
NGC5419  &  E   &  --  &  inc.  &  no  &  yes   \\
NGC5596  &  S0 &  yes &  inc.  &    no  &    --  \\
NGC5846  &  E   & inc.  &  inc.  &  no  &  --   \\
NGC6703  &  S0 &  --  & yes &  no  &  --   \\
NGC7077  &  E   &  --  & --  & yes &  --   \\
NGC7626  &  E   &  --  & inc.  &  no  &  yes  
\enddata
\label{tab:tab_agn}
\end{deluxetable}

\section{Colors}
\label{sec:colors}

Colors can be good and simple indicators of star formation activity in
galaxies. Therefore, in this section, we compare our previous results with
some typical colors. We compute a color-magnitude diagram (SDSS u-r
 color versus R magnitude) of galaxies for which we have SDSS data 
(79 E and 68 S0). In Fig.\ref{fig:11}, Elliptical and Lenticular galaxies 
are represented by red and green dots respectively, and as expected most galaxies 
($\sim$ 73\%) in our sample reside in the red sequence \citep{Strateva:01,Mignoli:09}. 
Among the remaining galaxies, 22\% (18 E and 15 S0) reside in the green valley 
and 5\% (2 E and 5 S0) in the blue cloud \citep{Strateva:01}.
The color-magnitude distribution of ETGs does not seem to depend on the galaxy
morphology apart from a statistically poor hint of dominance of S0 in the blue clouds.\\ 
In Fig.\ref{fig:11}, black squares outline a 160 $\mu$m MIPS detection, 
which indicates the presence of cold dust in galaxies. The 160 $\mu$m detections (92) 
are bound to brighter galaxies (R < -19.5) (74) or to galaxies in the green valley (14), 
this distribution is explained by a larger specific dust mass and sSFR in galaxies 
from the green valley.

\begin{figure}[h!t]
\hspace{-0.5cm}$\,$\includegraphics[width=9cm]{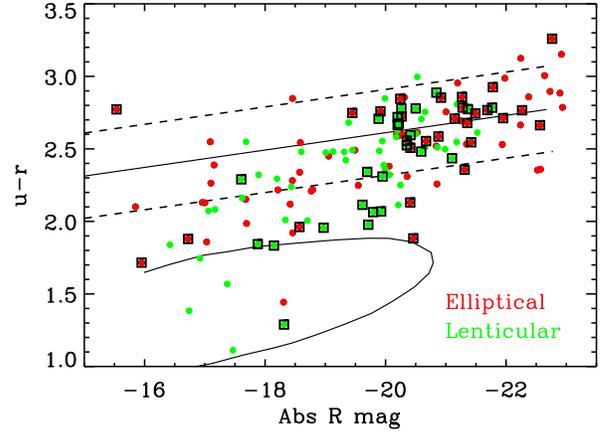}
\caption{Optical color (u-r) versus the absolute R magnitude computed using SDSS data 
of the 147 sources in our sample for which such data exist. Elliptical and Lenticular 
galaxies are represented by respectively red and green dots. Sources with a 160 $\mu$m
MIPS detection have been outlined with a black square. The solid straight line represents
the red sequence and the two dashed lines delimit the 3$\sigma$ confidence interval.
The solid curved line marks the position of the blue cloud. These area were taken from 
\cite{Temi:09b} and are not derived from the data points.
}
\label{fig:11}
\end{figure}

Fig.\,\ref{fig:12} shows the evolution of the dust-to-stellar mass ratio
 in logarithmic scale versus the (NUV-r) color. The dashed line corresponds 
to the best fit for our 29 sources with a secure SED fit and detected in the NUV. We find the
 same trend than \citep[e.g.,][]{Smith:12,Agius:13}, i.e. a decreasing M$_{dust}$/M$_{star}$ ratio 
 with an increasing (NUV-r) color. We color-code the stellar mass on each point of this 
plot with red (log(M$_{star})<$9.5), green (9.5 $<$ log(M$_{star})<$10.5), cyan 
(10.5 $<$ log(M$_{star})<$11) and blue dots (log(M$_{star})>$11). We see that the most 
massive galaxies (in stellar mass), i.e. galaxies with log(M$_{star})>$10.5 (cyan and 
blue dots), are very red ( (NUV-r) > 4). The most massive sources (blue dots) 
present a very low M$_{dust}$/M$_{star}$ ratio, which corresponds most preferentially 
to galaxies with low M$_{dust}$ and massive elliptical galaxies, as seen on 
Fig.\,\ref{fig:7}. The NUV-r color seems to correlate with
the dust mass only for bluer galaxies (NUV-r < 4.5), this result is consistent
with the relation between sSFR and M$_d$/M$_*$, and the slope (-0.57 $\pm$ 0.17) is compatible
with our slope in Fig. \ref{fig:7} and results from \cite{Rowlands:12} assuming a linear
relation between sSFR and NUV-r. The correlation is more clear for the lower mass
range of our sample (log(M$_*$/M$_\odot$) < 10.

\begin{figure}[ht!]
\hspace{-0.5cm}$\,$\includegraphics[width=9cm]{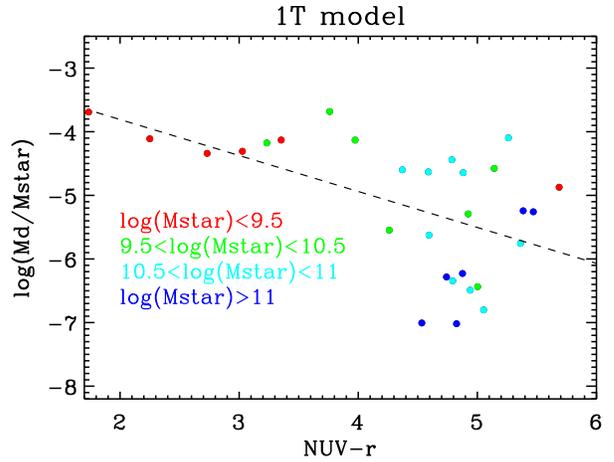}
\caption{ M$_{dust}$/M$_{star}$ versus (NUV-r) color. The dashed line correspond to the 
best linear fit for all sources detected in NUV and with a relative error on M$_{dust}$ $<$ 50\%. T
he red dots correspond to the least massive galaxies (in terms of stellar mass) of the sample (
log(M$_{star})<$9.5) and the most massive sources are represented by the blue dots.}
\label{fig:12}
\end{figure}

We test our sSFR estimated using CIGALEMC with the ratio F$_{24}$/F$_K$, which 
is in general a good indicator of specific star-formation when its value
is larger than 0.1 \cite{Temi:09b}. Fig. \ref{fig:13} shows a linear relation
between the sSFR versus and the 24 $\mu$m color. Most of the galaxies with
a large sSFR and 24 $\mu$m are S0 ($\sim$ 80\%). This confirms our previous
finding that S0 are more active in forming stars than E galaxies.\\
Three elliptical galaxies occupy a peculiar location in Fig. \ref{fig:13},
at log(sSFR)$\sim$-10 and F$_{24}$/F$_K\sim$0.05. These galaxies are 
IC3370, NGC4494 and NGC5077. They all host an AGN according to the Veron 
catalog\footnote{http://heasarc.gsfc.nasa.gov/W3Browse/all/veroncat.html} \citep{Veron:10}.
According to our SED fitting, their age$_{D4000}$, determined with the 4000 {\AA} 
 break, is also very low (age$_{D4000}$ < 1.5 Gyr), which would imply they had a recent episod of 
star formation.\\ IC3370 geometry has been found to be triaxial
\citep{Samurovic:05} and \cite{Jarvis:87} classified it as a peculiar S0pec
 with a box shape, and found some large isophotal twisting ($\Delta$P.A.$\simeq$ 25$\deg$)
in its bulge. \cite{Jarvis:87} also argue its peculiar shape could be due to the slow
merger of 2 massive disk galaxies following \cite{Binney:85} model.\\
NGC4494 has been found to have an unsually low X-ray flux compared to its B band
flux \citep{OSullivan:01b}, and this low ratio could be a hint for a recent merger
\citep{OSullivan:04} since in general it is found to be low for young early-type galaxies
\citep{Mackie:97,Sansom:00,OSullivan:01a}. We also found earlier that NGC4494 had unusually
low dust content for such high star formation rate.\\ NGC5077 is a LINER, with a H$\alpha$ bright
gaseous disk in its central area \citep{Bertola:91,Pizzella:97}. We have a hint that
2 out of the 3 outliers could be merger, which could explain their unusual 24 $\mu$m
color. However, there is also the possibility that our fit converge to a wrong solution
 for this 3 galaxies, since we do not have UV (GALEX) nor optical data (SDSS) for these galaxies.

\begin{figure}[h!t]
\hspace{-0.5cm}$\,$\includegraphics[width=9cm]{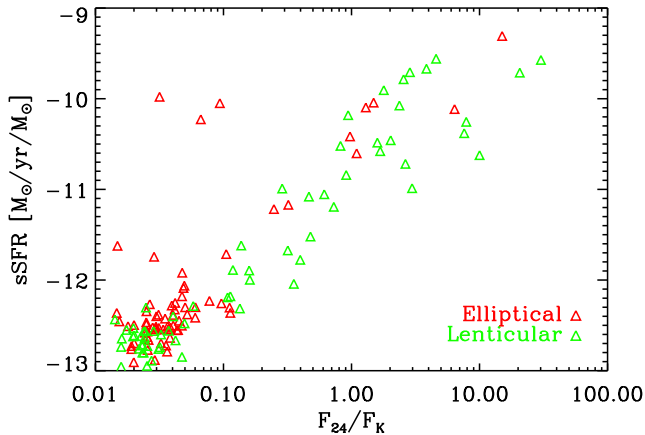}
\caption{Specific SFR estimated with the SED fitting versus the ratio F$_{24}$/F$_K$ for
our elliptical (red) and lenticular (green) galaxies. The sSFR estimate correlate well
with the 24 $\mu$m color.
}
\label{fig:13}
\end{figure}

Using GALEX and SDSS data, we compute the UV-optical color FUV-r on our sample where
data were available (72 E and 65 S0). Galaxies with infrared color F$_{24}$/F$_K$ less
than 0.1 are clustered around a FUV-r color of 7 and F$_{24}$/F$_K$ of 0.03 on figure \ref{fig:14}. 
Galaxies with larger infrared color are also bluer spreading in FUV-r from 6 to 2. Above a 0.1  
F$_{24}$/F$_K$ color, the logarithm of the infrared color linearly decreases as the
FUV-r color increases. Bluer galaxies with FUV-r < 6 and F$_{24}$/F$_K$ > 0.1  
are composed of 7 Elliptical and 16 Lenticular galaxies. It is another indication
that S0 galaxies are in proportion more active than Ellipticals and
show a clear agreement between infrared and UV star formation estimators.

\begin{figure}[h!t]
\hspace{-0.5cm}$\,$\includegraphics[width=9cm]{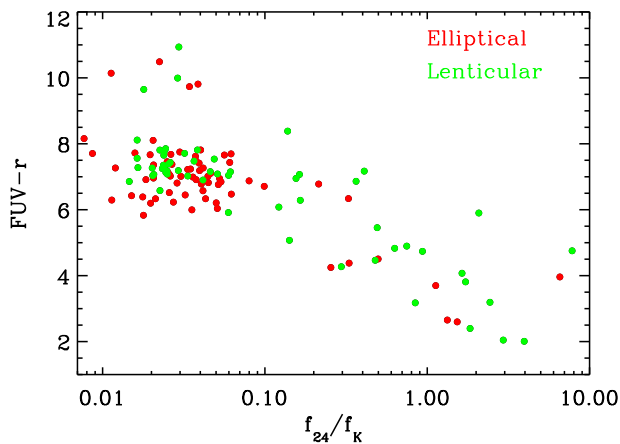}
\caption{Far-UV-R color calculated from GALEX and SDSS data as a function of the infrared
color F$_{24}$/f$_K$ , obtained from MIPS and 2MASS data. Elliptical and Lenticular galaxies 
are represented by respectively red and green dots.}
\label{fig:14}
\end{figure}

\section{Conclusions}
\label{sec:conclusion}

This paper analyzed a sample of 221 local ETGs to measure their star-formation activity and 
emphasize  differences between Elliptical and Lenticular galaxies. Indeed, most ETGs are 
red galaxies but studies from  past decades \citep[e.g.,][]{Faber:07,Schawinski:07b} implied 
an evolution from blue galaxies to red ones through a green phase possibly powered by an AGN. 
52\% of our sample are Elliptical and 48\% are S0. We performed  SED-fittings on our galaxies 
which produced constraints on their physical parameters, we concentrated on their star formation rate, dust luminosity, stellar and dust mass, and dust temperature. Using these parameters, we derived the following conclusions : 

\begin{itemize}
\item S0 have a larger specific star formation rate and larger dust luminosity when renormalized
by their stellar mass than elliptical galaxies. 
\item The average dust temperature of the sample is about 32.3\,K  fitting with a single temperature
 model, but it decreases to 22.4\,K when using a 2 temperature model. This difference most likely comes 
from the fact that our 2 temperature model requires at least one Herschel-SPIRE detection, and thus 
biases the selection towards colder galaxies. We did not find any difference between elliptical and
lenticular galaxy dust temperature.
\item The dust-to-stellar mass ratio is decreasing with the stellar mass for both elliptical and lenticular galaxies, and these relations are statistically indistiguishable.
\item The sSFR  does not evolve with the dust-to-stellar mass ratio for 
log(M$_{d}$/M$_{*}$)$<$-5 but steeply increases at higher dust-to-stellar mass ratio. 
Elliptical galaxies are mainly located at log(M$_{d}$/M$_{*}$)$<$-5 with no correlation between their 
sSFR and dust-to-stellar mass ratio, while SO galaxies are located at higher log(M$_{d}$/M$_{*}$), which correlated well with the specific star formation rate. 
\item According to our BPT and NIR diagrams,  a larger fraction ($\sim$12-19\%) of S0 galaxies show signs of AGN activity when compared to elliptical galaxies ($\sim$0-1\%). This observation is consistent with a scenario where blue 
star-forming galaxies evolve to a red elliptical passing through a S0 AGN active phase \citep{Schawinski:07b}.
\item far-infrared and UV colors are both in good agreement with our multi-frequency analysis
and with each other. All indicators point towards S0 being more active at forming stars on average
than ellipticals.
\end{itemize}

These results emphasize the differences between elliptical and lenticular galaxies to
possibly explain the transition phase between blue and red galaxies through the green valley. 
We will focus in the near future on the green and blue early-type galaxies, identified in this
paper, to have a better understanding of the star-formation transition phase, using  
GALEX, IRAC and MIPS data.

{\acknowledgements
We would like to thank our anonymous referee for his/her useful comments which
helped improve the original manuscript.
This publication makes use of data products from the Two Micron All Sky Survey, which is a joint project of the University of Massachusetts and the Infrared Processing and Analysis Center/California Institute of Technology, funded by the National Aeronautics and Space Administration and the National Science Foundation. This publication makes use of data from SDSS-III. Funding for SDSS-III has been provided by the Alfred P. Sloan Foundation, the Participating Institutions, the National Science Foundation, and the U.S. Department of Energy Office of Science. The SDSS-III web site is http://www.sdss3.org/. SDSS-III is managed by the Astrophysical Research Consortium for the Participating Institutions of the SDSS-III Collaboration including the University of Arizona, the Brazilian Participation Group, Brookhaven National Laboratory, University of Cambridge, Carnegie Mellon University, University of Florida, the French Participation Group, the German Participation Group, Harvard University, the Instituto de Astrofisica de Canarias, the Michigan State/Notre Dame/JINA Participation Group, Johns Hopkins University, Lawrence Berkeley National Laboratory, Max Planck Institute for Astrophysics, Max Planck Institute for Extraterrestrial Physics, New Mexico State University, New York University, Ohio State University, Pennsylvania State University, University of Portsmouth, Princeton University, the Spanish Participation Group, University of Tokyo, University of Utah, Vanderbilt University, University of Virginia, University of Washington, and Yale University. This work is based in part on observations made with the Spitzer Space Telescope, which is operated by the Jet Propulsion Laboratory, California Institute of Technology under a contract with NASA. This work is based in part on observations made with the NASA Galaxy Evolution Explorer. GALEX is operated for NASA by the California Institute of Technology under NASA contract NAS5-98034. 
}

\bibliography{biblio}

\appendix
{\LongTables
\input{table_flux1b}

\input{table_flux2b}

\input{table_flux4b}
\input{tableSpire}

\input{table_pars3}

\input{table_fdcem}

\input{table_fdcem2}

}
\end{document}

%% file: fig2.tex
\begingroup
  \makeatletter
  \providecommand\color[2][]{%
    \GenericError{(gnuplot) \space\space\space\@spaces}{%
      Package color not loaded in conjunction with
      terminal option `colourtext'%
    }{See the gnuplot documentation for explanation.%
    }{Either use 'blacktext' in gnuplot or load the package
      color.sty in LaTeX.}%
    \renewcommand\color[2][]{}%
  }%
  \providecommand\includegraphics[2][]{%
    \GenericError{(gnuplot) \space\space\space\@spaces}{%
      Package graphicx or graphics not loaded%
    }{See the gnuplot documentation for explanation.%
    }{The gnuplot epslatex terminal needs graphicx.sty or graphics.sty.}%
    \renewcommand\includegraphics[2][]{}%
  }%
  \providecommand\rotatebox[2]{#2}%
  \@ifundefined{ifGPcolor}{%
    \newif\ifGPcolor
    \GPcolortrue
  }{}%
  \@ifundefined{ifGPblacktext}{%
    \newif\ifGPblacktext
    \GPblacktexttrue
  }{}%
  \let\gplgaddtomacro\g@addto@macro
  \gdef\gplbacktext{}%
  \gdef\gplfronttext{}%
  \makeatother
  \ifGPblacktext
    \def\colorrgb#1{}%
    \def\colorgray#1{}%
  \else
    \ifGPcolor
      \def\colorrgb#1{\color[rgb]{#1}}%
      \def\colorgray#1{\color[gray]{#1}}%
      \expandafter\def\csname LTw\endcsname{\color{white}}%
      \expandafter\def\csname LTb\endcsname{\color{black}}%
      \expandafter\def\csname LTa\endcsname{\color{black}}%
      \expandafter\def\csname LT0\endcsname{\color[rgb]{1,0,0}}%
      \expandafter\def\csname LT1\endcsname{\color[rgb]{0,1,0}}%
      \expandafter\def\csname LT2\endcsname{\color[rgb]{0,0,1}}%
      \expandafter\def\csname LT3\endcsname{\color[rgb]{1,0,1}}%
      \expandafter\def\csname LT4\endcsname{\color[rgb]{0,1,1}}%
      \expandafter\def\csname LT5\endcsname{\color[rgb]{1,1,0}}%
      \expandafter\def\csname LT6\endcsname{\color[rgb]{0,0,0}}%
      \expandafter\def\csname LT7\endcsname{\color[rgb]{1,0.3,0}}%
      \expandafter\def\csname LT8\endcsname{\color[rgb]{0.5,0.5,0.5}}%
    \else
      \def\colorrgb#1{\color{black}}%
      \def\colorgray#1{\color[gray]{#1}}%
      \expandafter\def\csname LTw\endcsname{\color{white}}%
      \expandafter\def\csname LTb\endcsname{\color{black}}%
      \expandafter\def\csname LTa\endcsname{\color{black}}%
      \expandafter\def\csname LT0\endcsname{\color{black}}%
      \expandafter\def\csname LT1\endcsname{\color{black}}%
      \expandafter\def\csname LT2\endcsname{\color{black}}%
      \expandafter\def\csname LT3\endcsname{\color{black}}%
      \expandafter\def\csname LT4\endcsname{\color{black}}%
      \expandafter\def\csname LT5\endcsname{\color{black}}%
      \expandafter\def\csname LT6\endcsname{\color{black}}%
      \expandafter\def\csname LT7\endcsname{\color{black}}%
      \expandafter\def\csname LT8\endcsname{\color{black}}%
    \fi
  \fi
  \setlength{\unitlength}{0.0500bp}%
  \begin{picture}(11520.00,11520.00)%
    \gplgaddtomacro\gplbacktext{%
      \colorrgb{0.00,0.00,0.00}%
      \put(7836,717){\makebox(0,0){\strut{}log(L$_{bol}$) [L$_{\odot}$]}}%
    }%
    \gplgaddtomacro\gplfronttext{%
      \colorrgb{0.00,0.00,0.00}%
      \put(6992,1438){\makebox(0,0)[r]{\strut{}}}%
      \colorrgb{0.00,0.00,0.00}%
      \put(6992,1684){\makebox(0,0)[r]{\strut{}}}%
      \colorrgb{0.00,0.00,0.00}%
      \put(6992,1930){\makebox(0,0)[r]{\strut{}}}%
      \colorrgb{0.00,0.00,0.00}%
      \put(6992,2176){\makebox(0,0)[r]{\strut{}}}%
      \colorrgb{0.00,0.00,0.00}%
      \put(6992,2421){\makebox(0,0)[r]{\strut{}}}%
      \colorrgb{0.00,0.00,0.00}%
      \put(6992,2667){\makebox(0,0)[r]{\strut{}}}%
      \csname LTb\endcsname%
      \put(7351,1047){\makebox(0,0){\strut{}9.8}}%
      \put(7677,1047){\makebox(0,0){\strut{}9.9}}%
      \put(8002,1047){\makebox(0,0){\strut{}10}}%
      \put(8327,1047){\makebox(0,0){\strut{}10.1}}%
    }%
    \gplgaddtomacro\gplbacktext{%
      \colorrgb{0.00,0.00,0.00}%
      \put(5961,717){\makebox(0,0){\strut{}age$_{D4000}$ [Gyr]}}%
    }%
    \gplgaddtomacro\gplfronttext{%
      \colorrgb{0.00,0.00,0.00}%
      \put(5116,1438){\makebox(0,0)[r]{\strut{}}}%
      \colorrgb{0.00,0.00,0.00}%
      \put(5116,1684){\makebox(0,0)[r]{\strut{}}}%
      \colorrgb{0.00,0.00,0.00}%
      \put(5116,1930){\makebox(0,0)[r]{\strut{}}}%
      \colorrgb{0.00,0.00,0.00}%
      \put(5116,2176){\makebox(0,0)[r]{\strut{}}}%
      \colorrgb{0.00,0.00,0.00}%
      \put(5116,2421){\makebox(0,0)[r]{\strut{}}}%
      \colorrgb{0.00,0.00,0.00}%
      \put(5116,2667){\makebox(0,0)[r]{\strut{}}}%
      \csname LTb\endcsname%
      \put(5388,1047){\makebox(0,0){\strut{}8}}%
      \put(5833,1047){\makebox(0,0){\strut{}9}}%
      \put(6278,1047){\makebox(0,0){\strut{}10}}%
    }%
    \gplgaddtomacro\gplbacktext{%
    }%
    \gplgaddtomacro\gplfronttext{%
      \colorrgb{0.00,0.00,0.00}%
      \put(5116,3484){\makebox(0,0)[r]{\strut{}}}%
      \colorrgb{0.00,0.00,0.00}%
      \put(5116,3811){\makebox(0,0)[r]{\strut{}}}%
      \colorrgb{0.00,0.00,0.00}%
      \put(5116,4138){\makebox(0,0)[r]{\strut{}}}%
      \colorrgb{0.00,0.00,0.00}%
      \put(5116,4465){\makebox(0,0)[r]{\strut{}}}%
      \colorrgb{0.00,0.00,0.00}%
      \put(5388,3036){\makebox(0,0){\strut{}}}%
      \colorrgb{0.00,0.00,0.00}%
      \put(5833,3036){\makebox(0,0){\strut{}}}%
      \colorrgb{0.00,0.00,0.00}%
      \put(6278,3036){\makebox(0,0){\strut{}}}%
    }%
    \gplgaddtomacro\gplbacktext{%
      \colorrgb{0.00,0.00,0.00}%
      \put(4085,717){\makebox(0,0){\strut{}log(SFR) [M$_{\odot}$/yr]}}%
    }%
    \gplgaddtomacro\gplfronttext{%
      \colorrgb{0.00,0.00,0.00}%
      \put(3241,1438){\makebox(0,0)[r]{\strut{}}}%
      \colorrgb{0.00,0.00,0.00}%
      \put(3241,1684){\makebox(0,0)[r]{\strut{}}}%
      \colorrgb{0.00,0.00,0.00}%
      \put(3241,1930){\makebox(0,0)[r]{\strut{}}}%
      \colorrgb{0.00,0.00,0.00}%
      \put(3241,2176){\makebox(0,0)[r]{\strut{}}}%
      \colorrgb{0.00,0.00,0.00}%
      \put(3241,2421){\makebox(0,0)[r]{\strut{}}}%
      \colorrgb{0.00,0.00,0.00}%
      \put(3241,2667){\makebox(0,0)[r]{\strut{}}}%
      \csname LTb\endcsname%
      \put(3751,1047){\makebox(0,0){\strut{}-3}}%
      \put(4143,1047){\makebox(0,0){\strut{}-2.5}}%
      \put(4535,1047){\makebox(0,0){\strut{}-2}}%
    }%
    \gplgaddtomacro\gplbacktext{%
    }%
    \gplgaddtomacro\gplfronttext{%
      \colorrgb{0.00,0.00,0.00}%
      \put(3241,3484){\makebox(0,0)[r]{\strut{}}}%
      \colorrgb{0.00,0.00,0.00}%
      \put(3241,3811){\makebox(0,0)[r]{\strut{}}}%
      \colorrgb{0.00,0.00,0.00}%
      \put(3241,4138){\makebox(0,0)[r]{\strut{}}}%
      \colorrgb{0.00,0.00,0.00}%
      \put(3241,4465){\makebox(0,0)[r]{\strut{}}}%
      \colorrgb{0.00,0.00,0.00}%
      \put(3751,3036){\makebox(0,0){\strut{}}}%
      \colorrgb{0.00,0.00,0.00}%
      \put(4143,3036){\makebox(0,0){\strut{}}}%
      \colorrgb{0.00,0.00,0.00}%
      \put(4535,3036){\makebox(0,0){\strut{}}}%
    }%
    \gplgaddtomacro\gplbacktext{%
    }%
    \gplgaddtomacro\gplfronttext{%
      \colorrgb{0.00,0.00,0.00}%
      \put(3241,5387){\makebox(0,0)[r]{\strut{}}}%
      \colorrgb{0.00,0.00,0.00}%
      \put(3241,5610){\makebox(0,0)[r]{\strut{}}}%
      \colorrgb{0.00,0.00,0.00}%
      \put(3241,5833){\makebox(0,0)[r]{\strut{}}}%
      \colorrgb{0.00,0.00,0.00}%
      \put(3241,6056){\makebox(0,0)[r]{\strut{}}}%
      \colorrgb{0.00,0.00,0.00}%
      \put(3241,6279){\makebox(0,0)[r]{\strut{}}}%
      \colorrgb{0.00,0.00,0.00}%
      \put(3241,6502){\makebox(0,0)[r]{\strut{}}}%
      \colorrgb{0.00,0.00,0.00}%
      \put(3751,5026){\makebox(0,0){\strut{}}}%
      \colorrgb{0.00,0.00,0.00}%
      \put(4143,5026){\makebox(0,0){\strut{}}}%
      \colorrgb{0.00,0.00,0.00}%
      \put(4535,5026){\makebox(0,0){\strut{}}}%
    }%
    \gplgaddtomacro\gplbacktext{%
      \colorrgb{0.00,0.00,0.00}%
      \put(727,1982){\rotatebox{90}{\makebox(0,0){\strut{}log(L$_{dust}$) [L$_{\odot}$]}}}%
      \colorrgb{0.00,0.00,0.00}%
      \put(2210,717){\makebox(0,0){\strut{}log(M$_{*}$) [M$_{\odot}$]}}%
    }%
    \gplgaddtomacro\gplfronttext{%
      \csname LTb\endcsname%
      \put(1365,1438){\makebox(0,0)[r]{\strut{}5.5}}%
      \put(1365,1684){\makebox(0,0)[r]{\strut{}6}}%
      \put(1365,1930){\makebox(0,0)[r]{\strut{}6.5}}%
      \put(1365,2176){\makebox(0,0)[r]{\strut{}7}}%
      \put(1365,2421){\makebox(0,0)[r]{\strut{}7.5}}%
      \put(1365,2667){\makebox(0,0)[r]{\strut{}8}}%
      \put(1526,1047){\makebox(0,0){\strut{}10.2}}%
      \put(2129,1047){\makebox(0,0){\strut{}10.4}}%
      \put(2732,1047){\makebox(0,0){\strut{}10.6}}%
    }%
    \gplgaddtomacro\gplbacktext{%
      \colorrgb{0.00,0.00,0.00}%
      \put(595,3972){\rotatebox{90}{\makebox(0,0){\strut{}log(L$_{bol}$) [L$_{\odot}$]}}}%
    }%
    \gplgaddtomacro\gplfronttext{%
      \csname LTb\endcsname%
      \put(1365,3484){\makebox(0,0)[r]{\strut{}9.8}}%
      \put(1365,3811){\makebox(0,0)[r]{\strut{}9.9}}%
      \put(1365,4138){\makebox(0,0)[r]{\strut{}10}}%
      \put(1365,4465){\makebox(0,0)[r]{\strut{}10.1}}%
      \colorrgb{0.00,0.00,0.00}%
      \put(1526,3036){\makebox(0,0){\strut{}}}%
      \colorrgb{0.00,0.00,0.00}%
      \put(2129,3036){\makebox(0,0){\strut{}}}%
      \colorrgb{0.00,0.00,0.00}%
      \put(2732,3036){\makebox(0,0){\strut{}}}%
    }%
    \gplgaddtomacro\gplbacktext{%
      \colorrgb{0.00,0.00,0.00}%
      \put(595,5961){\rotatebox{90}{\makebox(0,0){\strut{}age$_{D4000}$ [Gyr]}}}%
    }%
    \gplgaddtomacro\gplfronttext{%
      \csname LTb\endcsname%
      \put(1365,5387){\makebox(0,0)[r]{\strut{}8}}%
      \put(1365,5610){\makebox(0,0)[r]{\strut{}8.5}}%
      \put(1365,5833){\makebox(0,0)[r]{\strut{}9}}%
      \put(1365,6056){\makebox(0,0)[r]{\strut{}9.5}}%
      \put(1365,6279){\makebox(0,0)[r]{\strut{}10}}%
      \put(1365,6502){\makebox(0,0)[r]{\strut{}10.5}}%
      \colorrgb{0.00,0.00,0.00}%
      \put(1526,5026){\makebox(0,0){\strut{}}}%
      \colorrgb{0.00,0.00,0.00}%
      \put(2129,5026){\makebox(0,0){\strut{}}}%
      \colorrgb{0.00,0.00,0.00}%
      \put(2732,5026){\makebox(0,0){\strut{}}}%
    }%
    \gplgaddtomacro\gplbacktext{%
      \colorrgb{0.00,0.00,0.00}%
      \put(595,7950){\rotatebox{90}{\makebox(0,0){\strut{}log(SFR) [M$_{\odot}$/yr]}}}%
    }%
    \gplgaddtomacro\gplfronttext{%
      \csname LTb\endcsname%
      \put(1365,7614){\makebox(0,0)[r]{\strut{}-3}}%
      \put(1365,8008){\makebox(0,0)[r]{\strut{}-2.5}}%
      \put(1365,8402){\makebox(0,0)[r]{\strut{}-2}}%
      \colorrgb{0.00,0.00,0.00}%
      \put(1526,7015){\makebox(0,0){\strut{}}}%
      \colorrgb{0.00,0.00,0.00}%
      \put(2129,7015){\makebox(0,0){\strut{}}}%
      \colorrgb{0.00,0.00,0.00}%
      \put(2732,7015){\makebox(0,0){\strut{}}}%
    }%
    \gplgaddtomacro\gplbacktext{%
      \csname LTb\endcsname%
      \put(9170,1047){\makebox(0,0){\strut{}5.5}}%
      \put(9660,1047){\makebox(0,0){\strut{}6.5}}%
      \put(10150,1047){\makebox(0,0){\strut{}7.5}}%
      \colorrgb{0.00,0.00,0.00}%
      \put(9712,717){\makebox(0,0){\strut{}log(L$_{dust}$) [L$_{\odot}$]}}%
    }%
    \gplgaddtomacro\gplfronttext{%
    }%
    \gplgaddtomacro\gplbacktext{%
      \csname LTb\endcsname%
      \put(7351,3036){\makebox(0,0){\strut{}9.8}}%
      \put(7677,3036){\makebox(0,0){\strut{}9.9}}%
      \put(8002,3036){\makebox(0,0){\strut{}10}}%
      \put(8327,3036){\makebox(0,0){\strut{}10.1}}%
    }%
    \gplgaddtomacro\gplfronttext{%
    }%
    \gplgaddtomacro\gplbacktext{%
      \csname LTb\endcsname%
      \put(5388,5026){\makebox(0,0){\strut{}8}}%
      \put(5833,5026){\makebox(0,0){\strut{}9}}%
      \put(6278,5026){\makebox(0,0){\strut{}10}}%
    }%
    \gplgaddtomacro\gplfronttext{%
    }%
    \gplgaddtomacro\gplbacktext{%
      \csname LTb\endcsname%
      \put(3751,7015){\makebox(0,0){\strut{}-3}}%
      \put(4143,7015){\makebox(0,0){\strut{}-2.5}}%
      \put(4535,7015){\makebox(0,0){\strut{}-2}}%
    }%
    \gplgaddtomacro\gplfronttext{%
    }%
    \gplgaddtomacro\gplbacktext{%
      \csname LTb\endcsname%
      \put(1526,9004){\makebox(0,0){\strut{}10.2}}%
      \put(2129,9004){\makebox(0,0){\strut{}10.4}}%
      \put(2732,9004){\makebox(0,0){\strut{}10.6}}%
    }%
    \gplgaddtomacro\gplfronttext{%
    }%
    \gplbacktext
    \put(0,0){\includegraphics{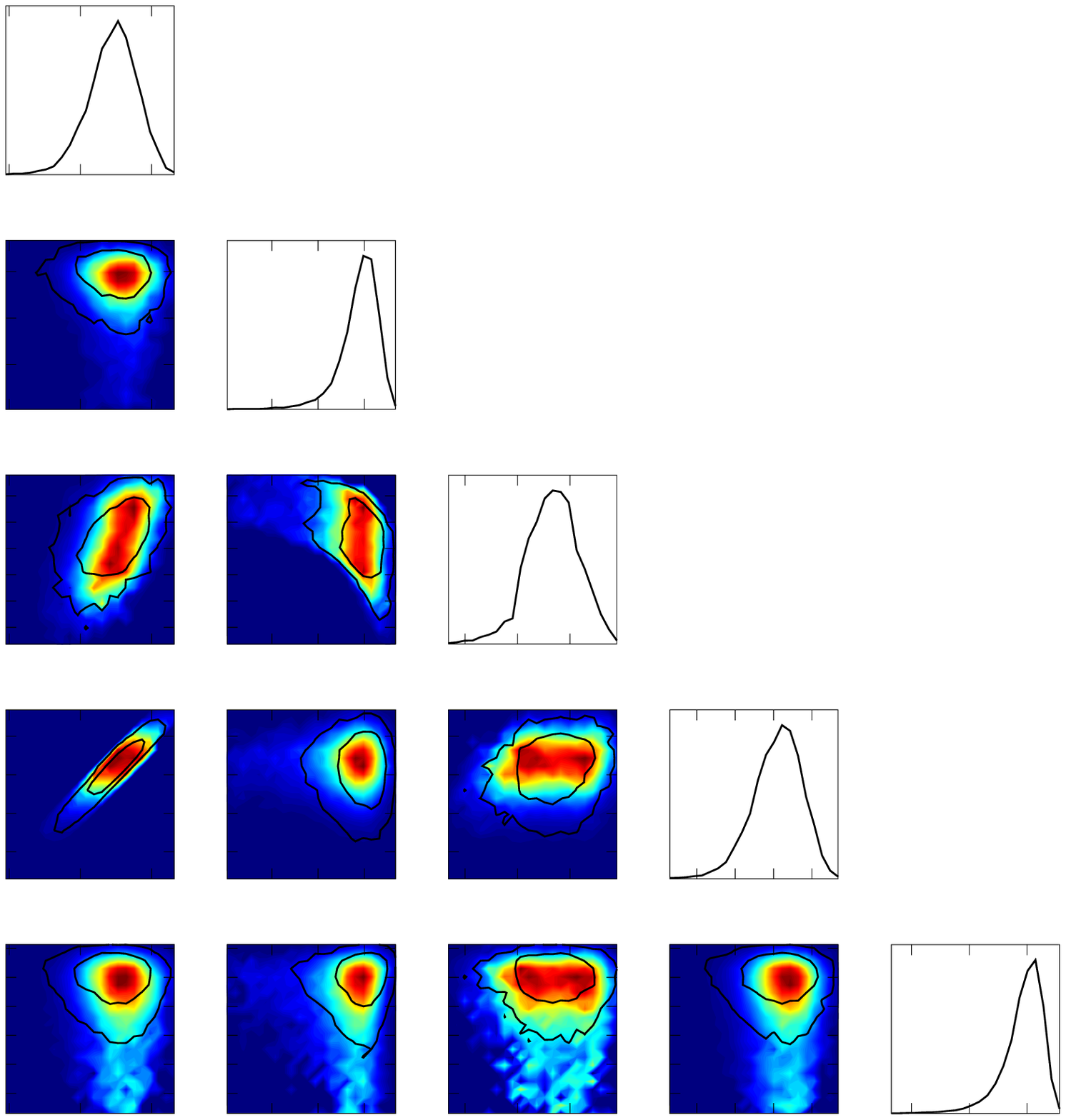}}%
    \gplfronttext
  \end{picture}%
\endgroup

%% file: table_flux1b.tex
\begin{deluxetable*}{llrrrrrrr}[h!]  
\tabletypesize{\footnotesize}  

\tablecaption{Fluxes (FUV to $z$ band, from GALEX and SDSS data) and redshifts used in this study, data were collected according to section \ref{sec:data}}
  
\label{tab:flux1} 
\tablehead{ 
\colhead{Name}  &  Redshift\tablenotemark{a} &  \colhead{$FUV$}  &  \colhead{$NUV$}  &  \colhead{$u$}  &  \colhead{$g$}  &  \colhead{$r$}  &  \colhead{$i$}  &  \colhead{$z$} \\
  &  &  \colhead{($\mu$Jy)}  &  \colhead{($\mu$Jy)} &  \colhead{(mJy)} &  \colhead{(mJy)} &  \colhead{(mJy)} &  \colhead{(mJy)} &  \colhead{(mJy)}
}    
\startdata  
Eso103-35 	 &  0.0133  & 	 - 	 & 	 - 	 & 	 - 	 & 	 - 	 & 	 - 	 & 	 - 	 & 	 - 	\\
																	  
Eso428-14 	 &  0.0057  & 	 - 	 & 	 - 	 & 	 - 	 & 	 - 	 & 	 - 	 & 	 - 	 & 	 - 	\\
																	  
Eso462-15 	 &  0.0191  & 	 76.33$\pm$8.11 	 & 	 351.30$\pm$17.22 	 & 	 - 	 & 	 - 	 & 	 - 	 & 	 - 	 & 	 - 	\\
																	  
Eso483-13 	 &  0.0030  & 	 1541.00$\pm$10.35 	 & 	 2364.00$\pm$10.02 	 & 	 - 	 & 	 - 	 & 	 - 	 & 	 - 	 & 	 - 	\\
																	  
IC0798 	 &  0.0047  & 	 16.75$\pm$2.97 	 & 	 133.10$\pm$4.56 	 & 	 1.24$\pm$0.02 	 & 	 4.71$\pm$0.01 	 & 	 8.58$\pm$0.02 	 & 	 11.97$\pm$0.03 	 & 	 14.31$\pm$0.10 	\\
																	  
IC1144 	 &  0.0403  & 	 1.58$\pm$0.46 	 & 	 184.20$\pm$3.95 	 & 	 1.50$\pm$0.02 	 & 	 6.99$\pm$0.08 	 & 	 13.24$\pm$0.15 	 & 	 17.94$\pm$0.21 	 & 	 22.13$\pm$0.24 	\\
																	  
IC1459 	 &  0.0060  & 	 781.10$\pm$8.39 	 & 	 3327.00$\pm$15.47 	 & 	 - 	 & 	 - 	 & 	 - 	 & 	 - 	 & 	 - 	\\
																	  
IC1639 	 &  0.0179  & 	 33.79$\pm$1.23 	 & 	 189.70$\pm$2.34 	 & 	 2.03$\pm$0.01 	 & 	 8.81$\pm$0.02 	 & 	 16.26$\pm$0.03 	 & 	 22.03$\pm$0.05 	 & 	 26.80$\pm$0.07 	\\
\enddata  
\tablenotetext{a}{For a few galaxies (mainly very nearby galaxies), the redshift was calculated using the distance from NED.}  
  
\end{deluxetable*}

%% file: table_flux2b.tex
\begin{deluxetable*}{lrrrrrrr}[h!]
\tabletypesize{\footnotesize}

\label{tab:flux2}
\tablecaption{Fluxes (2MASS J to K band and Spitzer IRAC 3.6 to 8 $\mu$m) used in this study, data were collected according to section \ref{sec:data}}

\tablehead{
\colhead{Name} & \colhead{$J$} & \colhead{$H$} & \colhead{$Ks$} & \colhead{$i3.6$} & \colhead{$i4.5$} & \colhead{$i5.8$} & \colhead{$i8.0$} \\
 & \colhead{(mJy)}& \colhead{(mJy)}& \colhead{(mJy)}& \colhead{(mJy)}& \colhead{(mJy)}& \colhead{(mJy)}& \colhead{(mJy)}
}

\startdata

Eso103-35 	&	 46.14$\pm$1.02 	&	 58.98$\pm$1.58 	&	 50.70$\pm$1.64	&	 - 	&	 - 	&	 - 	&	 - 	\\
															
Eso428-14 	&	 197.20$\pm$2.36 	&	 252.50$\pm$3.02 	&	 213.30$\pm$4.52	&	 121.60$\pm$6.08 	&	 81.35$\pm$4.07 	&	 119.10$\pm$5.96 	&	 249.40$\pm$12.47 	\\
															
Eso462-15 	&	 197.60$\pm$2.37 	&	 242.50$\pm$3.80 	&	 205.20$\pm$3.78	&	 138.20$\pm$6.91 	&	 73.53$\pm$3.68 	&	 54.14$\pm$2.71 	&	 22.39$\pm$1.12 	\\
															
Eso483-13 	&	 35.33$\pm$1.53 	&	 36.47$\pm$2.39 	&	 31.67$\pm$2.86	&	 13.53$\pm$0.68 	&	 8.80$\pm$0.44 	&	 7.26$\pm$0.36 	&	 10.22$\pm$0.51 	\\
															
IC0798 	&	 17.54$\pm$0.63 	&	 21.24$\pm$1.12 	&	 15.21$\pm$1.05	&	 - 	&	 - 	&	 - 	&	 - 	\\
															
IC1144 	&	 33.96$\pm$0.88 	&	 43.21$\pm$1.43 	&	 37.20$\pm$1.27	&	 20.56$\pm$1.03 	&	 11.88$\pm$0.59 	&	 7.67$\pm$0.38 	&	 5.87$\pm$0.29 	\\
															
IC1459 	&	 1373.00$\pm$24.03 	&	 1573.00$\pm$30.43 	&	 1301.00$\pm$29.95	&	 591.90$\pm$29.60 	&	 290.40$\pm$14.52 	&	 156.00$\pm$7.80 	&	 140.10$\pm$7.01 	\\
															
IC1639 	&	 38.35$\pm$0.60 	&	 46.17$\pm$0.94 	&	 38.42$\pm$1.03	&	 19.10$\pm$0.96 	&	 11.38$\pm$0.57 	&	 8.12$\pm$0.41 	&	 5.48$\pm$0.27 	\\
\enddata

\end{deluxetable*}

%% file: table_flux4b.tex
\begin{deluxetable*}{lrrrrrrrr}[h!]
\tabletypesize{\footnotesize}

\label{tab:flux3}
\tablecaption{Fluxes (24 $\mu$m to 160 $\mu$m Spitzer MIPS, 12$\mu$m to 100 $\mu$m IRAS)  used in this study, data were collected according to section \ref{sec:data}}

\tablehead{
\colhead{Name} & \colhead{$24_{\mu m}$} & \colhead{$70_{\mu m}$} & \colhead{$160_{\mu m}$}& \colhead{$I12_{\mu m}$}& \colhead{$I25_{\mu m}$}& \colhead{$I60_{\mu m}$}& \colhead{$I100_{\mu m}$} & \colhead{H$\alpha$}\\
 & \colhead{(mJy)}& \colhead{(mJy)}& \colhead{(mJy)}& \colhead{(mJy)}& \colhead{(mJy)}& \colhead{(mJy)}& \colhead{(mJy)} & \colhead{(erg/s/cm$^2$)}\\
}

\startdata

Eso103-35 	 & 	 1693.0$\pm$81.0 	 & 	 1774.0$\pm$112.0 	 & 	 - 	 & 	 360$\pm$73 	 & 	 40$\pm$59 	 & 	 410$\pm$321 	 & 15170$\pm$3886 & 	 - 	\\
														
Eso428-14 	 & 	 1448.0$\pm$80.0 	 & 	 4714.0$\pm$389.0 	 & 	 - 	 & 	 20$\pm$37 	 & 	 340$\pm$70 	 & 	 430$\pm$80 	 & 70$\pm$348 & 	 - 	\\
														
Eso462-15 	 & 	 6.1$\pm$0.7 	 & 	 0.0$\pm$19.4 	 & 	 0.0$\pm$26.8 	 & 	 810$\pm$21871 	 & 	 - 	 & 	 1790$\pm$1432 	 & 4960$\pm$774 & 	 - 	\\
														
Eso483-13 	 & 	 29.8$\pm$3.8 	 & 	 417.0$\pm$23.0 	 & 	 579.0$\pm$69.0 	 & 	 40$\pm$41 	 & 	 - 	 & 	 150$\pm$57 	 & 1310$\pm$228 & 	 - 	\\
														
IC0798 	 & 	 0.6$\pm$0.2 	 & 	 - 	 & 	 - 	 & 	 20$\pm$34 	 & 	 120$\pm$68 	 & 	 160$\pm$43 	 & 470$\pm$191 & 16.05$\pm$3.72\\
														
IC1144 	 & 	 1.4$\pm$0.3 	 & 	 - 	 & 	 - 	 & 	 - 	 & 	 120$\pm$35 	 & 	 160$\pm$103 	 & - & 	 - 	\\
														
IC1459 	 & 	 60.9$\pm$4.3 	 & 	 542.0$\pm$32.0 	 & 	 627.0$\pm$72.0 	 & 	 - 	 & 	 - 	 & 	 - 	 & - & 	 - 	\\
														
IC1639 	 & 	 3.7$\pm$0.4 	 & 	 - 	 & 	 - 	 & 	 20$\pm$35 	 & 	 160$\pm$50 	 & 	 280$\pm$44 	 & 1210$\pm$158 & 49.57$\pm$10.38\\
\enddata
\end{deluxetable*}

%% file: tableSpire.tex
\begin{deluxetable*}{cccc}
\tabletypesize{\scriptsize}
\tablecaption{Herschel SPIRE fluxes used in this work.}

\label{tab:spire}

\tablehead{
\colhead{Name} & \colhead{$250_{\mu m}$} & \colhead{$350_{\mu m}$} & \colhead{$500_{\mu m}$}\\
\colhead{} & \colhead{(mJy)} & \colhead{(mJy)} & \colhead{(mJy)}
}
\startdata
IC3032
& 0.0
$\pm$3.2
& 0.0
$\pm$2.9
& 0.0
$\pm$2.1
\\
IC3101
& 0.0
$\pm$3.4
& 0.0
$\pm$2.8
& 0.0
$\pm$2.1
\\
IC3328
& 0.0
$\pm$3.8
& 0.0
$\pm$3.6
& 0.6
$\pm$2.6
\\

IC3381
& 1.1
$\pm$5.4
& 0.6
$\pm$4.8
& 4.0
$\pm$3.9
\\

IC3501
& 0.0
$\pm$3.7
& 0.0
$\pm$3.3
& 0.0
$\pm$2.6
\\

IC5063
& 1723.8
$\pm$258.7
& 848.1
$\pm$134.0
& 279.4
$\pm$53.1
\\

NGC0584
& 0.8
$\pm$13.9
& 6.2
$\pm$13.4
& 2.7
$\pm$9.7
\\

NGC0855
& 788.2
$\pm$118.3
& 351.6
$\pm$53.0
& 139.9
$\pm$36.7
\\
\enddata
\end{deluxetable*}

%% file: table_pars3.tex
\begin{deluxetable*}{lrrrrrrr}
\tabletypesize{\scriptsize}
\label{tab:pars}
\tablecaption{Parameters fitted by CIGALEMC using GALEX, SDSS, 2MASS, IRAC and MIPS datawhen available.}

\tablehead{\colhead{Name} & \colhead{Log($f_{ysp}$)} & \colhead{Log($M_{gal}$)} & \colhead{Log($M_\star$)} & \colhead{Log(SFR)} & \colhead{$t_{D4000}$} & \colhead{Log($L_{bol}$)} & \colhead{Log($L_d$)}\\
\colhead{} &\colhead{} & \colhead{($M_\sun$)} & \colhead{($M_\sun$)} & \colhead{($M_\sun/yr$)} & \colhead{(Gyr)} & \colhead{($L_\sun$)} & \colhead{($L_\sun$)}}
\startdata
Eso462-15& -3.25$\pm$0.83& 11.65$\pm$0.89& 11.51$\pm$0.89& -1.01$\pm$0.28& 9.27$\pm$2.53& 11.09$\pm$0.61& 8.66$\pm$0.40\\
Eso483-13& -1.23$\pm$0.57& 9.10$\pm$0.18& 8.97$\pm$0.18& -1.22$\pm$0.09& 0.59$\pm$0.20& 9.00$\pm$0.07& 8.04$\pm$0.15\\
IC0798& -4.53$\pm$0.93& 9.63$\pm$0.06& 9.49$\pm$0.06& -3.19$\pm$0.64& 9.09$\pm$0.94& 9.02$\pm$0.06& 6.41$\pm$0.55\\
IC1144& -0.46$\pm$0.45& 7.77$\pm$0.96& 7.64$\pm$0.96& -1.94$\pm$0.94& 0.37$\pm$0.38& 8.35$\pm$0.84& 8.24$\pm$0.88\\
IC1459& -2.52$\pm$1.05& 11.04$\pm$1.05& 10.90$\pm$1.04& -0.92$\pm$0.22& 7.39$\pm$3.60& 10.61$\pm$0.66& 8.87$\pm$0.15\\
IC1639& -3.01$\pm$0.42& 11.09$\pm$0.06& 10.94$\pm$0.06& -1.32$\pm$0.23& 8.91$\pm$0.91& 10.47$\pm$0.05& 8.53$\pm$0.38\\
IC3328& -0.65$\pm$0.55& 6.44$\pm$0.33& 6.31$\pm$0.33& -3.55$\pm$0.40& 0.53$\pm$0.65& 6.62$\pm$0.29& 6.48$\pm$0.33\\
IC3370& -1.16$\pm$0.72& 11.40$\pm$0.20& 11.27$\pm$0.19& 1.04$\pm$0.44& 1.16$\pm$1.52& 11.32$\pm$0.16& 9.32$\pm$0.20\\
\enddata
\end{deluxetable*}

%% file: table_fdcem.tex
\begin{deluxetable*}{lrrrr}
\tabletypesize{\scriptsize}
\tablecaption{Parameters fitted with single temperature model and emissivity $\beta$=1.5 on FIR data.}
\tablewidth{0pt}

\tablehead{
\colhead{Name} & \colhead{Log($L_d$)} & \colhead{T} & \colhead{Log($M_d$)} & \colhead{$\chi^2 / ndf$}\\
\colhead{} &\colhead{($L_\sun$)} & \colhead{(K)} & \colhead{($M_\sun$)} & \colhead{}
}
\startdata
Eso103-35 & 10.57$\pm$0.05 & 88.33$\pm$4.40 & 4.90$\pm$0.06 & 10.10\\
Eso428-14 & 9.76$\pm$0.02 & 54.14$\pm$3.61 & 5.22$\pm$0.14 & 1.05\\
Eso483-13 & 8.12$\pm$0.02 & 31.45$\pm$1.40 & 4.87$\pm$0.11 & 50.00\\
IC1459 & 8.75$\pm$0.02 & 31.76$\pm$1.31 & 5.47$\pm$0.10 & 1.18\\
IC3032 & 6.23$\pm$0.82 & 61.10$\pm$21.56 & 2.71$\pm$2.20 & 0.06\\
IC3101 & 6.30$\pm$0.87 & 60.61$\pm$21.96 & 2.95$\pm$2.27 & 0.05\\
IC3328 & 6.28$\pm$0.79 & 61.56$\pm$21.29 & 2.66$\pm$2.40 & 3.12\\
IC3370 & 9.43$\pm$0.02 & 30.07$\pm$0.93 & 6.28$\pm$0.08 & 67.35\\
\enddata
\label{tab:td1}
\end{deluxetable*}

%% file: table_fdcem2.tex
\begin{deluxetable*}{lrrrrrr}
\tabletypesize{\scriptsize}

\tablecaption{Parameters fitted by a 2 temperature model with an emissivity $\beta$=2 using FIR data.}
\tablewidth{0pt}

\tablehead{
\colhead{Name} & \colhead{Log($L_d$)} & \colhead{$\alpha$} & \colhead{T$_d^1$} & \colhead{T$_d^2$}& \colhead{Log($M_d$)} & \colhead{$\chi^2$/ndf}\\
\colhead{} &\colhead{($L_\sun$)} & \colhead{} & \colhead{(K)} & \colhead{(K)} &\colhead{($M_\sun$)} & \colhead{}
}
\startdata
IC3328 & 6.45$\pm$0.81 & 0.50$\pm$0.29 & 27.48$\pm$10.13 & 71.11$\pm$14.39 & 4.75$\pm$2.80 & 9.31\\
IC5063 & 10.63$\pm$0.07 & 1.00$\pm$0.00 & 16.53$\pm$1.48 & 70.67$\pm$5.36 & 8.63$\pm$0.35 & 2.15\\
NGC0584 & 8.34$\pm$0.21 & 0.46$\pm$0.27 & 26.86$\pm$10.03 & 79.91$\pm$11.67 & 6.11$\pm$1.43 & 8.44\\
NGC0855 & 8.36$\pm$0.07 & 0.99$\pm$0.01 & 22.61$\pm$1.12 & 57.17$\pm$7.27 & 5.48$\pm$0.13 & 2.52\\
NGC1266 & 10.56$\pm$0.03 & 0.99$\pm$0.00 & 27.60$\pm$0.59 & 70.13$\pm$3.06 & 7.15$\pm$0.06 & 12.83\\
NGC1316 & 9.61$\pm$0.07 & 1.00$\pm$0.00 & 24.11$\pm$0.65 & 55.17$\pm$8.73 & 6.56$\pm$0.13 & 5.55\\
NGC1377 & 10.17$\pm$0.04 & 0.98$\pm$0.00 & 29.83$\pm$1.07 & 72.82$\pm$4.15 & 6.55$\pm$0.09 & 2.46\\
NGC1386 & 9.34$\pm$0.04 & 0.99$\pm$0.00 & 24.18$\pm$0.64 & 62.96$\pm$4.13 & 6.27$\pm$0.07 & 0.40\\
\enddata
\label{tab:td2}
\end{deluxetable*}